\documentclass[prd,twocolumn,superscriptaddress]{revtex4}

\usepackage{graphicx}

\usepackage{amsfonts}
\usepackage{amsmath}
\usepackage{amssymb}
\usepackage{bm}

\usepackage{float}
\usepackage{subfig}
\usepackage{array}
\usepackage{multirow}

\begin{document}
%\setpagewiselinenumbers
%\linenumbers

\captionsetup[figure]{font=small,justification=centerlast,singlelinecheck=false}
\captionsetup[table]{font=small,justification=centerlast,singlelinecheck=false}

\newcommand{\costheta}{$\cos\theta_{\pi^0}$}
\newcommand{\ppin}{$p_{\pi^0}$}
\newcommand{\dcostheta}{$d\cos\theta_{\pi^0}$}
\newcommand{\dppin}{$d\,p_{\pi^0}$}
\newcommand{\bv}[1]{\mathbf{#1}}

\newcommand{\nue}{$\nu_e$}
\newcommand{\nuebar}{$\bar{\nu}_e$}
\newcommand{\numu}{$\nu_\mu$}
\newcommand{\numubar}{$\bar{\nu}_\mu$}
\newcommand{\ncpi}{$\text{NC }1\pi^0$}
\newcommand{\piz}{$\pi^0$}

\title{Measurement of \numu{} and \numubar{} induced neutral current single $\pi^0$ production cross sections on mineral oil at $E_\nu\sim\mathcal{O}(1\text{ GeV})$}

\affiliation{University of Alabama; Tuscaloosa, AL 35487}
\affiliation{Bucknell University; Lewisburg, PA 17837}
\affiliation{University of Cincinnati; Cincinnati, OH 45221}
\affiliation{University of Colorado; Boulder, CO 80309}
\affiliation{Columbia University; New York, NY 10027}
\affiliation{Embry Riddle Aeronautical University; Prescott, AZ 86301}
\affiliation{Fermi National Accelerator Laboratory; Batavia, IL 60510}
\affiliation{University of Florida; Gainesville, FL 32611}
\affiliation{University of Illinois; Urbana, IL 61801}
\affiliation{Indiana University; Bloomington, IN 47405}
\affiliation{Instituto de Ciencias Nucleares, Universidad Nacional Aut\'onoma de M\'exico, D.F. 04510, M\'exico}
\affiliation{Los Alamos National Laboratory; Los Alamos, NM 87545}
\affiliation{Louisiana State University; Baton Rouge, LA 70803}
\affiliation{Massachusetts Institute of Technology; Cambridge, MA 02139}
\affiliation{University of Michigan; Ann Arbor, MI 48109}
\affiliation{Princeton University; Princeton, NJ 08544}
\affiliation{Saint Mary's University of Minnesota; Winona, MN 55987}
\affiliation{Virginia Polytechnic Institute \& State University; Blacksburg, VA 24061}
\affiliation{Yale University; New Haven, CT 06520}
\author{A.~A. Aguilar-Arevalo}
\affiliation{Instituto de Ciencias Nucleares, Universidad Nacional Aut\'onoma de M\'exico, D.F. 04510, M\'exico}
\author{C.~E.~Anderson}
\affiliation{Yale University; New Haven, CT 06520}
\author{A.~O.~Bazarko}
\affiliation{Princeton University; Princeton, NJ 08544}
\author{S.~J.~Brice}
\affiliation{Fermi National Accelerator Laboratory; Batavia, IL 60510}
\author{B.~C.~Brown}
\affiliation{Fermi National Accelerator Laboratory; Batavia, IL 60510}
\author{L.~Bugel}
\affiliation{Columbia University; New York, NY 10027}
\author{J.~Cao}
\affiliation{University of Michigan; Ann Arbor, MI 48109}
\author{L.~Coney}
\affiliation{Columbia University; New York, NY 10027}
\author{J.~M.~Conrad}
\affiliation{Massachusetts Institute of Technology; Cambridge, MA 02139}
\author{D.~C.~Cox}
\affiliation{Indiana University; Bloomington, IN 47405}
\author{A.~Curioni}
\affiliation{Yale University; New Haven, CT 06520}
\author{Z.~Djurcic}
\affiliation{Columbia University; New York, NY 10027}
\author{D.~A.~Finley}
\affiliation{Fermi National Accelerator Laboratory; Batavia, IL 60510}
\author{B.~T.~Fleming}
\affiliation{Yale University; New Haven, CT 06520}
\author{R.~Ford}
\affiliation{Fermi National Accelerator Laboratory; Batavia, IL 60510}
\author{F.~G.~Garcia}
\affiliation{Fermi National Accelerator Laboratory; Batavia, IL 60510}
\author{G.~T.~Garvey}
\affiliation{Los Alamos National Laboratory; Los Alamos, NM 87545}
\author{J.~Gonzales}
\affiliation{Los Alamos National Laboratory; Los Alamos, NM 87545}
\author{J.~Grange}
\affiliation{University of Florida; Gainesville, FL 32611}
\author{C.~Green}
\affiliation{Fermi National Accelerator Laboratory; Batavia, IL 60510}
\affiliation{Los Alamos National Laboratory; Los Alamos, NM 87545}
\author{J.~A.~Green}
\affiliation{Indiana University; Bloomington, IN 47405}
\affiliation{Los Alamos National Laboratory; Los Alamos, NM 87545}
\author{T.~L.~Hart}
\affiliation{University of Colorado; Boulder, CO 80309}
\author{E.~Hawker}
\affiliation{University of Cincinnati; Cincinnati, OH 45221}
\affiliation{Los Alamos National Laboratory; Los Alamos, NM 87545}
\author{R.~Imlay}
\affiliation{Louisiana State University; Baton Rouge, LA 70803}
\author{R.~A. ~Johnson}
\affiliation{University of Cincinnati; Cincinnati, OH 45221}
\author{G.~Karagiorgi}
\affiliation{Massachusetts Institute of Technology; Cambridge, MA 02139}
\author{P.~Kasper}
\affiliation{Fermi National Accelerator Laboratory; Batavia, IL 60510}
\author{T.~Katori}
\affiliation{Indiana University; Bloomington, IN 47405}
\affiliation{Massachusetts Institute of Technology; Cambridge, MA 02139}
\author{T.~Kobilarcik}
\affiliation{Fermi National Accelerator Laboratory; Batavia, IL 60510}
\author{I.~Kourbanis}
\affiliation{Fermi National Accelerator Laboratory; Batavia, IL 60510}
\author{S.~Koutsoliotas}
\affiliation{Bucknell University; Lewisburg, PA 17837}
\author{E.~M.~Laird}
\affiliation{Princeton University; Princeton, NJ 08544}
\author{S.~K.~Linden}
\affiliation{Yale University; New Haven, CT 06520}
\author{J.~M.~Link}
\affiliation{Virginia Polytechnic Institute \& State University; Blacksburg, VA 24061}
\author{Y.~Liu}
\affiliation{University of Alabama; Tuscaloosa, AL 35487}
\author{Y.~Liu}
\affiliation{University of Michigan; Ann Arbor, MI 48109}
\author{W.~C.~Louis}
\affiliation{Los Alamos National Laboratory; Los Alamos, NM 87545}
\author{K.~B.~M.~Mahn}
\affiliation{Columbia University; New York, NY 10027}
\author{W.~Marsh}
\affiliation{Fermi National Accelerator Laboratory; Batavia, IL 60510}
\author{C.~Mauger}
\affiliation{Los Alamos National Laboratory; Los Alamos, NM 87545}
\author{V.~T.~McGary}
\affiliation{Massachusetts Institute of Technology; Cambridge, MA 02139}
\author{G.~McGregor}
\affiliation{Los Alamos National Laboratory; Los Alamos, NM 87545}
\author{W.~Metcalf}
\affiliation{Louisiana State University; Baton Rouge, LA 70803}
\author{P.~D.~Meyers}
\affiliation{Princeton University; Princeton, NJ 08544}
\author{F.~Mills}
\affiliation{Fermi National Accelerator Laboratory; Batavia, IL 60510}
\author{G.~B.~Mills}
\affiliation{Los Alamos National Laboratory; Los Alamos, NM 87545}
\author{J.~Monroe}
\affiliation{Columbia University; New York, NY 10027}
\author{C.~D.~Moore}
\affiliation{Fermi National Accelerator Laboratory; Batavia, IL 60510}
\author{J.~Mousseau}
\affiliation{University of Florida; Gainesville, FL 32611}
\author{R.~H.~Nelson}
\affiliation{University of Colorado; Boulder, CO 80309}
\author{P.~Nienaber}
\affiliation{Saint Mary's University of Minnesota; Winona, MN 55987}
\author{J.~A.~Nowak}
\affiliation{Louisiana State University; Baton Rouge, LA 70803}
\author{B.~Osmanov}
\affiliation{University of Florida; Gainesville, FL 32611}
\author{S.~Ouedraogo}
\affiliation{Louisiana State University; Baton Rouge, LA 70803}
\author{R.~B.~Patterson}
\affiliation{Princeton University; Princeton, NJ 08544}
\author{Z.~Pavlovic}
\affiliation{Los Alamos National Laboratory; Los Alamos, NM 87545}
\author{D.~Perevalov}
\affiliation{University of Alabama; Tuscaloosa, AL 35487}
\author{C.~C.~Polly}
\affiliation{Fermi National Accelerator Laboratory; Batavia, IL 60510}
\author{E.~Prebys}
\affiliation{Fermi National Accelerator Laboratory; Batavia, IL 60510}
\author{J.~L.~Raaf}
\affiliation{University of Cincinnati; Cincinnati, OH 45221}
\author{H.~Ray}
\affiliation{University of Florida; Gainesville, FL 32611}
\affiliation{Los Alamos National Laboratory; Los Alamos, NM 87545}
\author{B.~P.~Roe}
\affiliation{University of Michigan; Ann Arbor, MI 48109}
\author{A.~D.~Russell}
\affiliation{Fermi National Accelerator Laboratory; Batavia, IL 60510}
\author{V.~Sandberg}
\affiliation{Los Alamos National Laboratory; Los Alamos, NM 87545}
\author{R.~Schirato}
\affiliation{Los Alamos National Laboratory; Los Alamos, NM 87545}
\author{D.~Schmitz}
\affiliation{Columbia University; New York, NY 10027}
\author{M.~H.~Shaevitz}
\affiliation{Columbia University; New York, NY 10027}
\author{F.~C.~Shoemaker\footnote{deceased}}
\affiliation{Princeton University; Princeton, NJ 08544}
\author{D.~Smith}
\affiliation{Embry Riddle Aeronautical University; Prescott, AZ 86301}
\author{M.~Soderberg}
\affiliation{Yale University; New Haven, CT 06520}
\author{M.~Sorel\footnote{Present address: IFIC, Universidad de Valencia and CSIC, Valencia 46071, Spain}}
\affiliation{Columbia University; New York, NY 10027}
\author{P.~Spentzouris}
\affiliation{Fermi National Accelerator Laboratory; Batavia, IL 60510}
\author{J.~Spitz}
\affiliation{Yale University; New Haven, CT 06520}
\author{I.~Stancu}
\affiliation{University of Alabama; Tuscaloosa, AL 35487}
\author{R.~J.~Stefanski}
\affiliation{Fermi National Accelerator Laboratory; Batavia, IL 60510}
\author{M.~Sung}
\affiliation{Louisiana State University; Baton Rouge, LA 70803}
\author{H.~A.~Tanaka}
\affiliation{Princeton University; Princeton, NJ 08544}
\author{R.~Tayloe}
\affiliation{Indiana University; Bloomington, IN 47405}
\author{M.~Tzanov}
\affiliation{University of Colorado; Boulder, CO 80309}
\author{R.~G.~Van~de~Water}
\affiliation{Los Alamos National Laboratory; Los Alamos, NM 87545}
\author{M.~O.~Wascko\footnote{Present address: Imperial College; London SW7 2AZ, United Kingdom}}
\affiliation{Louisiana State University; Baton Rouge, LA 70803}
\author{D.~H.~White}
\affiliation{Los Alamos National Laboratory; Los Alamos, NM 87545}
\author{M.~J.~Wilking}
\affiliation{University of Colorado; Boulder, CO 80309}
\author{H.~J.~Yang}
\affiliation{University of Michigan; Ann Arbor, MI 48109}
\author{G.~P.~Zeller}
\affiliation{Fermi National Accelerator Laboratory; Batavia, IL 60510}
\author{E.~D.~Zimmerman}
\affiliation{University of Colorado; Boulder, CO 80309}
\collaboration{The MiniBooNE Collaboration}\noaffiliation

\begin{abstract}
MiniBooNE reports the first absolute cross sections for neutral current single $\pi^0$ production on $\text{CH}_2$ induced by neutrino and antineutrino interactions measured from the largest sets of NC $\pi^0$ events collected to date.  The principal result consists of differential cross sections measured as functions of $\pi^0$ momentum and $\pi^0$ angle averaged over the neutrino flux at MiniBooNE.  We find total cross sections of $(4.76\pm0.05_{stat}\pm0.76_{sys})\times10^{-40}\text{ cm}^2/\text{nucleon}$ at a mean energy of $\langle E_\nu\rangle=808\text{ MeV}$ and $(1.48\pm0.05_{stat}\pm0.23_{sys})\times10^{-40}\text{ cm}^2/\text{nucleon}$ at a mean energy of $\langle E_\nu\rangle=664\text{ MeV}$ for \numu{} and \numubar{} induced production, respectively.  In addition, we have included measurements of the neutrino and antineutrino total cross sections for incoherent exclusive \ncpi{} production corrected for the effects of final state interactions to compare to prior results.
\end{abstract}

\maketitle

\section{\label{sec:intro}Introduction}
Neutral current neutrino interactions producing a single $\pi^0$ (\ncpi{}) constitute a substantial background for experiments searching for $\nu_\mu\rightarrow\nu_e$ oscillations. \ncpi{} events are prone to mimicking single electrons---the signature sought in such \nue{}-appearance searches---because one of the two photons from the $\pi^0$ decay may escape detection.  
In MiniBooNE, \ncpi{} production poses one of the largest backgrounds: it is second only to events induced by intrinsic $\nu_e$ in the beam\cite{OSCPAPER}.  As such, absolute measurements of \ncpi{} production at energies of $\mathcal{O}(1\text{ GeV})$ are crucial to constraining this background, especially as it applies to future long-baseline experiments.

A measurement of \ncpi{} production can also be used to test and refine models of single $\pi^0$ production, which vary widely in their predictions at these energies\cite{RSCOH,ALVAREZRUSO,AMAROPRD,RSRES,HERNANDEZ,BERGER,PASCHOS,LEITNERINCOH,LEITNERCOH,NAKAMURA,GERSHTEIN,KOMACHENKO,MARTINI,KELKAR,BELKOV,SINGH}.  These models categorize exclusive \ncpi{} production on nuclei by final state as either \emph{coherent} or \emph{incoherent}.  Production leaving the nuclear target in the ground state is defined as coherent, otherwise it is defined as incoherent.  Prior measurements of \ncpi{} production were typically limited in scope, having addressed incoherent and coherent production separately, and suffered from low statistics.  The earliest results were total cross sections measured as ratios normalized to various charged current pion production channels\cite{ANL1,ANL2,BNL,AP,GGM}.  Later, studies of absolute \ncpi{} production were performed.  Absolute measurements of incoherent \ncpi{} production were reported by Aachen-Padova\cite{FAISSNER} (albeit in a footnote) and in a more recent reanalysis of Gargamelle data\cite{GGMHAWKER}, both at neutrino energies near 2 GeV.  The distinct signature of coherent \ncpi{} production---a forward emitted $\pi^0$ and a target left in its ground state---permits absolute measurements of coherent \ncpi{} production.  These measurements were carried out under a variety of circumstances\cite{FAISSNER,GGMISIKSAL,SKAT,CHARM,NOMAD}.  While measurements regarding such \emph{exclusive} production are valuable, the total yield of \ncpi{} production is often more important to modern-day neutrino oscillation experiments.  To address this need, \emph{inclusive} \ncpi{} and NC $\pi^0$ measurements, reported as flux-averaged cross section ratios relative to CC production, have been recently performed by K2K\cite{K2K} and SciBooNE\cite{SCIBOONE}, respectively.  Collectively, prior experiments have recorded a few thousand neutrino and a few hundred antineutrino \ncpi{} interactions.

In this paper, MiniBooNE reports the first measurements of absolute inclusive \ncpi{} cross sections (not normalized as ratios) for both neutrino and antineutrino scattering.  We define signal \ncpi{} events to be NC interactions wherein only one $\pi^0$ and no additional meson exits the target nucleus (no requirement on the number or identity of outgoing nucleons is made).  This definition is consistent with that used at K2K\cite{K2K}.  It is specifically chosen because final state interactions (FSI) dramatically alter the experimentally observed products of the original neutrino interaction on a nuclear target, but are not well understood. As particles in the final state transit the nucleus, they can scatter, be absorbed, or undergo charge exchange.  The observation of \ncpi{} interactions in an experiment will be depleted by the effects of absorption and charge exchange ($\pi^0\text{p}\rightarrow\pi^+\text{n}$, $\pi^0\text{n}\rightarrow\pi^-\text{p}$); however, it can also be enhanced by additional channels entering the sample if a $\pi^0$ is produced via FSI (\textit{e.g.} $\pi^+\text{n}\rightarrow\pi^0\text{p}$,  $\pi^-\text{p}\rightarrow\pi^0\text{n}$, or $\pi^0$ production from nucleon rescattering).  Ultimately, it is this observed rate of $\pi^0$ production, regardless of the initial interaction, that is relevant to neutrino oscillation experiments operating on nuclear targets.  Hence, the definition of our signal, one constructed in terms of the observed final state, directly addresses the requirements for $\nu_\mu\rightarrow\nu_e$ oscillation experiments.  At the same time, the inclusivity of the definition reduces the dependence of the measurement on the assumed models of FSI and single $\pi^0$ production.  Hereafter, we use ``\ncpi{}'' to refer to this inclusive definition unless explicitly stated otherwise.  Under this definition and in a calculated effort to reduce model dependence, we present the first absolute differential and total cross sections for \numu{} and \numubar{} induced \ncpi{} production; the interactions occurred on $\textrm{CH}_2$. Since the neutrino energy cannot be measured for each interaction, the cross sections are necessarily averaged over the neutrino flux at MiniBooNE.  Specifically, we have measured cross sections as a function of $\pi^0$ momentum (\ppin{}) and $\pi^0$ angle relative to the interacting neutrino (\costheta{}).  Together, these measurements can yield important information on FSI effects, which are a strong function of $\pi^0$ momentum, and the production mechanism (coherent versus incoherent), which is a strong function of $\pi^0$ angle.

\section{\label{sec:exp}The Experiment}
MiniBooNE receives neutrinos from the Booster Neutrino Beam at Fermilab. 8 GeV protons extracted from the Booster synchrotron are delivered to a beryllium target; neutrinos result from the decays of secondary mesons produced by interactions in the target.  The target is housed in a magnetic horn which focuses charged mesons of a selected sign and defocuses mesons of the opposite sign. A beam which is predominately composed of either neutrinos or antineutrinos can be produced by choosing the polarity of the horn current.  In neutrino mode, \numu{} with a mean energy of 808 MeV comprise 93.6\% of the flux and contamination from \numubar{}, \nue{}, and \nuebar{} comprise 5.86\%, 0.52\%, and 0.05\% of the flux, respectively.  Wrong-sign\footnote{Wrong-sign contamination refers to antineutrinos in the neutrino mode beam and neutrinos in the antineutrino mode beam} (WS) contamination impacts the antineutrino mode flux to a greater degree. In antineutrino mode, \numubar{} with a mean energy of 664 MeV comprise 83.73\% of the flux and contamination from \numu{}, \nue{}, and \nuebar{} comprise 15.71\%, 0.2\%, and 0.4\% of the flux, respectively\cite{FLUXPAPER}.

The detector\cite{DETECTOR} consists of a 12.2 m diameter spherical vessel filled with 818 tons of undoped mineral oil situated 541 m from the target.  The containment vessel is segmented by an optical barrier into a 5.75 m radius inner tank region and an additional 0.35 m veto region.  The surface of the inner tank is instrumented with 1280 8-inch photomultiplier tubes (PMTs), which provide 11.3\% photocathode coverage.  The tank PMTs capture the pattern of light generated by charged products of neutrino interactions.  Particles above Cherenkov threshold emit directional light conically about the particle track which produces a ring on the tank surface.  Isotropic scintillation light emitted by certain constituents of the mineral oil is also detected by the PMTs.  The veto region, which is instrumented with 240 PMTs, is used to detect light due to particles entering or exiting the detector.

Neutrino interactions in MiniBooNE are simulated using the v3 \textsc{Nuance} event generator\cite{NUANCE} coupled to a GEANT3-based\cite{GEANT} detector Monte Carlo.  Single $\pi^0$ production is predicted according to the models of Rein and Sehgal (R-S)\cite{RSRES,RSCOH} as implemented in \textsc{Nuance} with two exceptions.  First, we modify \textsc{Nuance} to incorporate nonisotropic $\Delta$ decays.  Second, the relative contribution of coherent and incoherent exclusive \ncpi{} production is further adjusted using a prior measurement\cite{PI0PLB}: coherent pion production is reduced by 35\% and incoherent is increased a corresponding 5\% to preserve total $\pi^0$ production.  The FSI model in \textsc{Nuance} accounts for the rescattering of all hadrons during nuclear transit; the pion absorption factor described in the R-S model of coherent pion production is omitted in lieu of the \textsc{Nuance} FSI model.  In all, we predict 94\% of observed \ncpi{} production to involve the production of a $\pi^0$ at the neutrino interaction vertex; the fraction rises to 97\% in antineutrino mode. A breakdown of the composition of \ncpi{} production by exclusive interaction channel is listed in Table~\ref{tab:signal}.  The R-S models predict a smaller incoherent pion production cross section for antineutrinos than for neutrinos, but similar coherent pion production cross sections for both.  As a result, the Monte Carlo predicts that the fraction of \ncpi{} production that is coherent pion production is larger in antineutrino mode than in neutrino mode.  In principle, this effect makes antineutrino scattering more sensitive to the coherent pion production mode.  The mineral oil target, which consists largely of long alkanes and cycloalkanes, is simulated as $\textrm{CH}_2$ in \textsc{Nuance}.  21\% of \ncpi{} production is predicted to occur on free nucleons (hydrogen).  This fraction is greater than the fraction of nucleons in $\textrm{CH}_2$ (14.3\%) belonging to H because nuclear effects (predominately pion absorption) diminish the cross section on carbon.
\begin{table}
\begin{ruledtabular}
\begin{tabular}{l@{\hspace{1em}}c@{\hspace{.5em}}c@{\hspace{1em}}l@{\hspace{1em}}c@{\hspace{.5em}}c}
Channel&$ \bm{\nu}  $&$ \bar{\bm{\nu} } $&Channel&$ \bm{\nu}  $&$ \bar{\bm{\nu} } $\\
\hline\\[-8pt]
NC $1\pi^0$ & 94\% & 97\% & NC Elastic & 2\% & $<1$\%\\
{\footnotesize\quad\textit{Incoherent}\cite{RSRES,NUANCE}} & {\footnotesize 77\%} & {\footnotesize 59\%} & Multi-$\pi$ & $<1$\% & $<1$\%\\
{\footnotesize\quad\textit{Coherent}\cite{RSCOH,NUANCE}} & {\footnotesize 17\%} & {\footnotesize 38\%} & DIS & $<1$\% & $<1$\%\\
NC $\pi^\pm$ & 2\% & 2\% & $\text{K},\rho,\eta$ Prod. & $<1$\% & $<1$\%\\
\end{tabular}
\end{ruledtabular}
\caption{\label{tab:signal}Predicted fractional composition of \ncpi{} signal events in neutrino and antineutrino modes broken down according to exclusive channel at the neutrino interaction vertex.}
\end{table}

\section{\label{sec:selection}Selection And Reconstruction}
Before events are reconstructed, a series of simple  cuts are made. Events are decomposed into sets of PMT hits clustered in time (subevents).  Selected \ncpi{} candidates are required to have (1) only one subevent and that subevent is coincident with the 1.6$\mu$s neutrino beam pulse.  Multiple subevents arise principally from muon decays---a signature of charged current events or $\pi^\pm$ production.  Further cuts require that the single subevent possess (2) fewer than 6 PMT hits in the veto region and (3) greater than 200 PMT hits in the tank region.  The veto hits requirement removes uncontained events as well as events with particles entering the detector during the beam pulse. The tank hits requirement reduces the contamination from NC elastic events and eliminates events containing a decay electron from a cosmic muon entering the tank before the beam.  

After the preliminary cuts, the remaining events are reconstructed in order to measure kinematic variables and perform particle identification.  The reconstruction algorithm takes the form of a track-based, least negative-log-likelihood fit performed under various particle hypotheses\cite{RECONSTRUCTION}.  Four hypotheses are used in this analysis: an electron ($e$) hypothesis, a muon ($\mu$) hypothesis, a two-photon ($\gamma\gamma$) hypothesis, and a pion ($\pi^0$) hypothesis.  The electron and muon fits are single track fits parametrized by vertex position $(x,y,z,t)$, direction $(\theta,\phi)$, and energy $(E)$.  The probability of the charge and time of each PMT hit resulting from a given track configuration can be estimated using an optical model including predictions for Cherenkov and scintillation light emission profiles for the outgoing lepton and a description of light propagation in the detector. The optical model is informed by \textit{in situ} measurements.  For each event, the negative-log-likelihood of the prediction compared to data is minimized over the space of track configurations.  The muon and electron hypotheses differ most significantly in the predicted topology of their associated Cherenkov rings.  Rings from electrons are blurred by multiple scattering and electromagnetic showers whereas muons, with straighter tracks and no associated showering, project sharp rings onto the surface of the detector.  The two-photon hypothesis is a two-track fit.  Conceptually, the two tracks represent the two photons from a \piz{} decay.  In practice, each track is treated using the electron hypothesis since photons resemble electrons in the detector.  The two tracks share a common vertex and are parametrized by direction and energy as in the one-track fit and are each also parametrized by the photon conversion length.  The $\pi^0$ hypothesis is enforced by constraining the photon-photon invariant mass $m_{\gamma\gamma}$ to the \piz{} mass in the two-photon fit.
\begin{figure}
\includegraphics[width=\columnwidth]{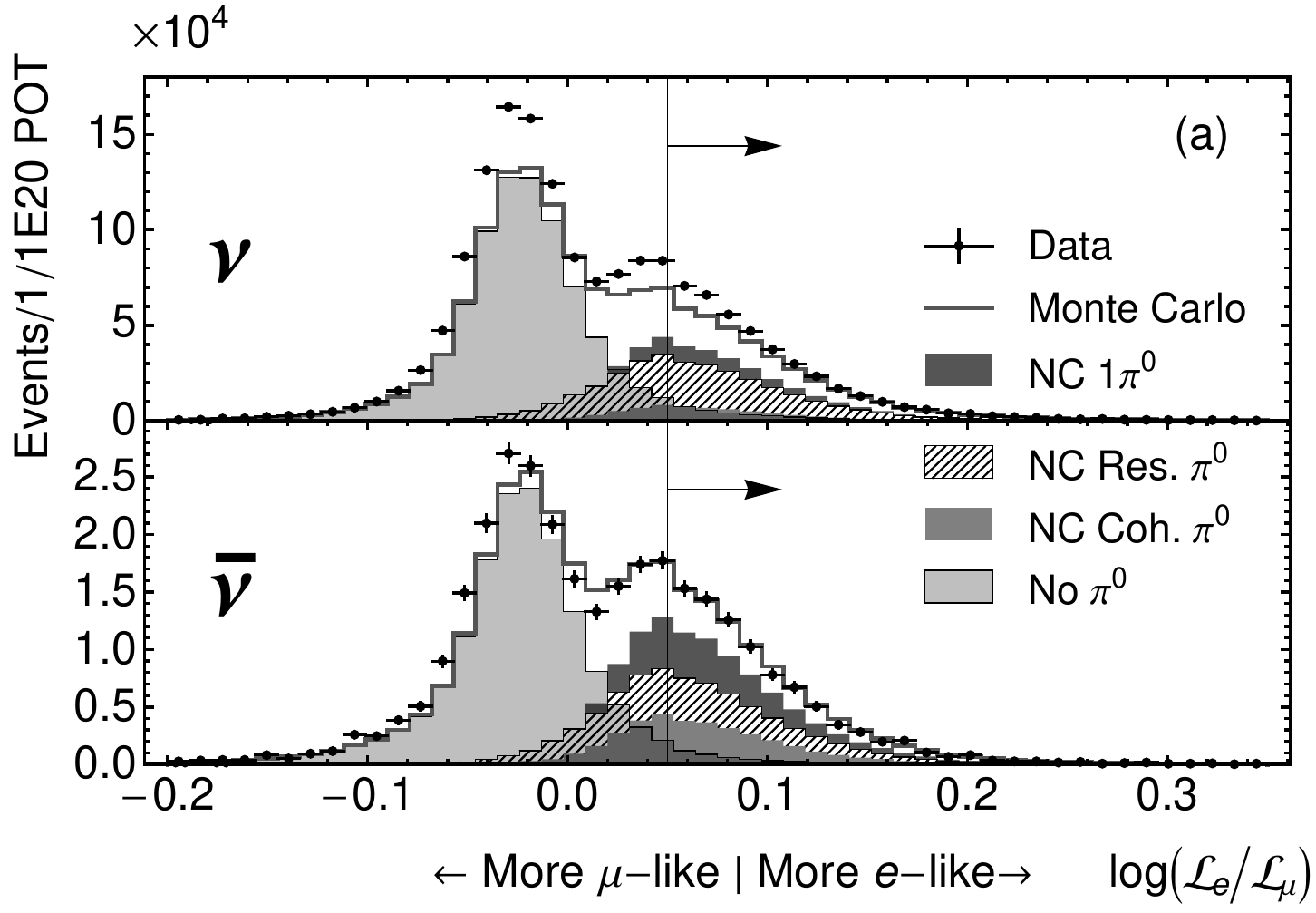}
\includegraphics[width=\columnwidth]{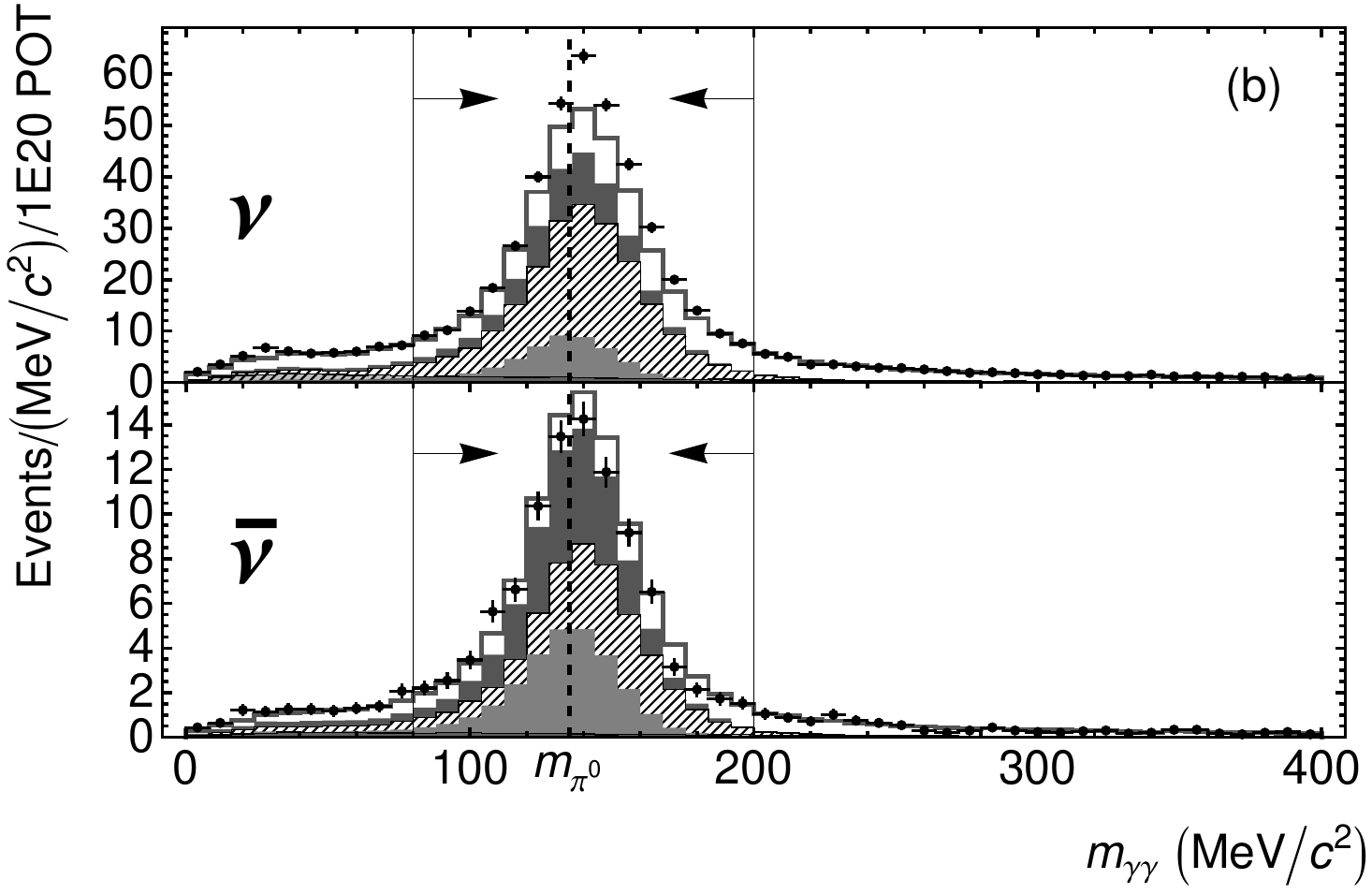}
\caption{\label{fig:piddists}(a) Distribution of the difference between the $e$ log-likelihood and the $\mu$ log-likelihood for events passing cuts (1)-(4) described in the text for neutrino mode running (top) and antineutrino mode running (bottom). Monte Carlo is depicted by a dark-gray line and data by black dots. Both data and Monte Carlo are absolutely normalized to $10^{20}$ POT. Error bars are statistical only.  Also shown are the contributions from events containing no $\pi^0$ in the detector (translucent light-gray fill), signal NC $1\pi^0$ production (dark-gray fill), and  incoherent (hatched fill) and coherent (gray fill) exclusive NC $1\pi^0$ production according to identification at the neutrino interaction vertex. Candidate \ncpi{} events are selected in the region indicated by the arrows. (b) Distribution of the reconstructed $\gamma$-$\gamma$ invariant mass for events passing cuts (1)-(6) described in the text.  The dashed vertical line marks the expected $\pi^0$ mass.}
\end{figure}
Reconstructed variables are used to further refine the \ncpi{} sample.  We require interaction vertices of candidates to (4) be within a 500 cm-radius fiducial volume according to the electron fit.  Candidates must favor the electron likelihood over the muon likelihood: more precisely, we require (5) $\log(\mathcal{L}_e/\mathcal{L}_\mu)>0.05$.  The distribution of this difference appears in Fig.~\ref{fig:piddists}.  The separation between events with and without a $\pi^0$ is evident.  Candidates must then favor the pion likelihood over the electron likelihood: (6) $\log(\mathcal{L}_e/\mathcal{L}_\pi)<0$.  Finally, we require that (7) the invariant mass extracted from the two-photon fit reside in the interval [80,200] $\text{MeV}/c^2$.  Figure~\ref{fig:piddists} includes the invariant mass distribution; a distinct peak around the $\pi^0$ mass of 134.97 $\text{MeV}/c^2$ is visible.  Only a miniscule number of events in the mass peak is predicted to contain no $\pi^0$s. A summary of the effect of each cut on the predicted purity and efficiency of each sample appears in Table~\ref{tab:cut}.

With $6.46\times10^{20}$ protons-on-target (POT) collected in neutrino mode running, 21375 events pass the selection requirements.  In antineutrino mode running, 2789 events pass selection requirements with $3.68\times10^{20}$ POT collected.  The Monte Carlo underestimates the number of events passing the cuts in neutrino mode by $10.9(8)_{stat}\%$ and overestimates it in antineutrino mode by $5(2)_{stat}\%$.  In each running mode, the sample collected is the largest set of NC $1\pi^0$ events recorded to date.  These samples exceed the total of all samples collected by previous experiments by roughly an order of magnitude.
\begin{table}
\begin{ruledtabular}
\begin{tabular}{>{$}l<{$}c@{\hspace{.5em}}cc@{\hspace{1em}}c@{\hspace{.5em}}c}
\multirow{2}{8em}{\parbox[c]{8em}{\large Cut}}&\multicolumn{2}{c}{Purity (w/\parbox[c]{3.7em}{\tiny Wrong\\Sign Signal})}&&\multicolumn{2}{c}{Efficiency}\\
\cline{2-6}\\[-8pt]
&$ \bm{\nu}  $&$ \bar{\bm{\nu} } $&$  $&$ \bm{\nu}  $&$ \bar{\bm{\nu} } $\\
\hline\\[-8pt]
 \text{None} & 5\% (5\%) & 4\% (6\%) &  & 100\% & 100\% \\ 
 \text{(1) }\text{1 Subevent} & 9\% (10\%) & 7\% (11\%) &  & 78\% & 78\% \\ 
 \text{(2) }N_{\text{Veto}} & 12\% (12\%) & 11\% (15\%) &  & 65\% & 67\% \\ 
 \text{(3) }N_{\text{Tank}} & 28\% (29\%) & 27\% (38\%) &  & 64\% & 65\% \\ 
 \text{(4) }R_e & 27\% (27\%) & 26\% (36\%) &  & 63\% & 62\% \\ 
 \text{(5) }\log \left(\mathcal{L}_e/\mathcal{L_{\mu }}\right) & 60\% (62\%) & 50\% (71\%) &  & 41\% & 40\% \\ 
 \text{(6) }\log \left(\mathcal{L}_e/\mathcal{L_{\pi }}\right) & 61\% (63\%) & 50\% (71\%) &  & 40\% & 39\% \\ 
 \text{(7) }m_{\gamma \gamma } & 73\% (75\%) & 58\% (82\%) &  & 36\% & 36\%\\ 
\end{tabular}
\end{ruledtabular}
\caption{\label{tab:cut}Predicted purity of the \ncpi{} sample and \ncpi{} selection efficiency in neutrino and antineutrino modes after each cut described in the text.  Purity including wrong-sign induced signal sources is presented parenthetically.}
\end{table}

\section{\label{sec:analysis}Analysis And Results}
\begin{figure*}
\includegraphics[width=\textwidth]{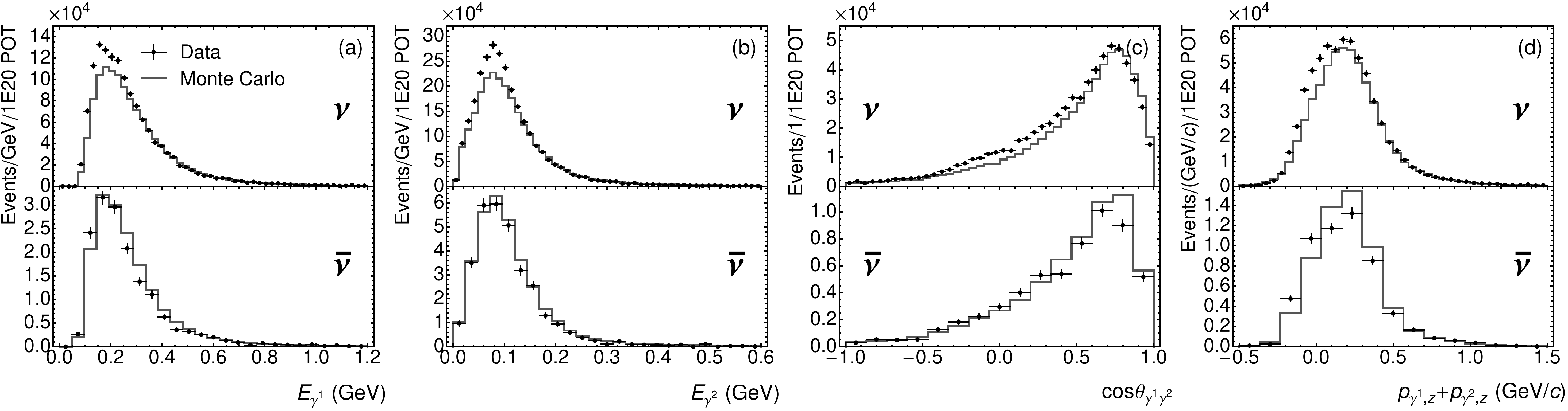}
\caption{\label{fig:photonkinematics}(a) The distribution of the reconstructed energy of the more energetic $\gamma$ from the $\pi^0$ decay in \ncpi{} candidates from Monte Carlo (dark-gray line) and data (black dots).  Results from neutrino mode running appear on the top and antineutrino mode running on the bottom. Error bars are statistical only and distributions are absolutely normalized to $10^{20}$ POT. (b) The reconstructed energy of the less energetic $\gamma$. (c) The reconstructed opening angle between the two photons. (d) The reconstructed total momentum in the beam direction.}
\end{figure*}
\begin{figure}
\includegraphics[width=\columnwidth]{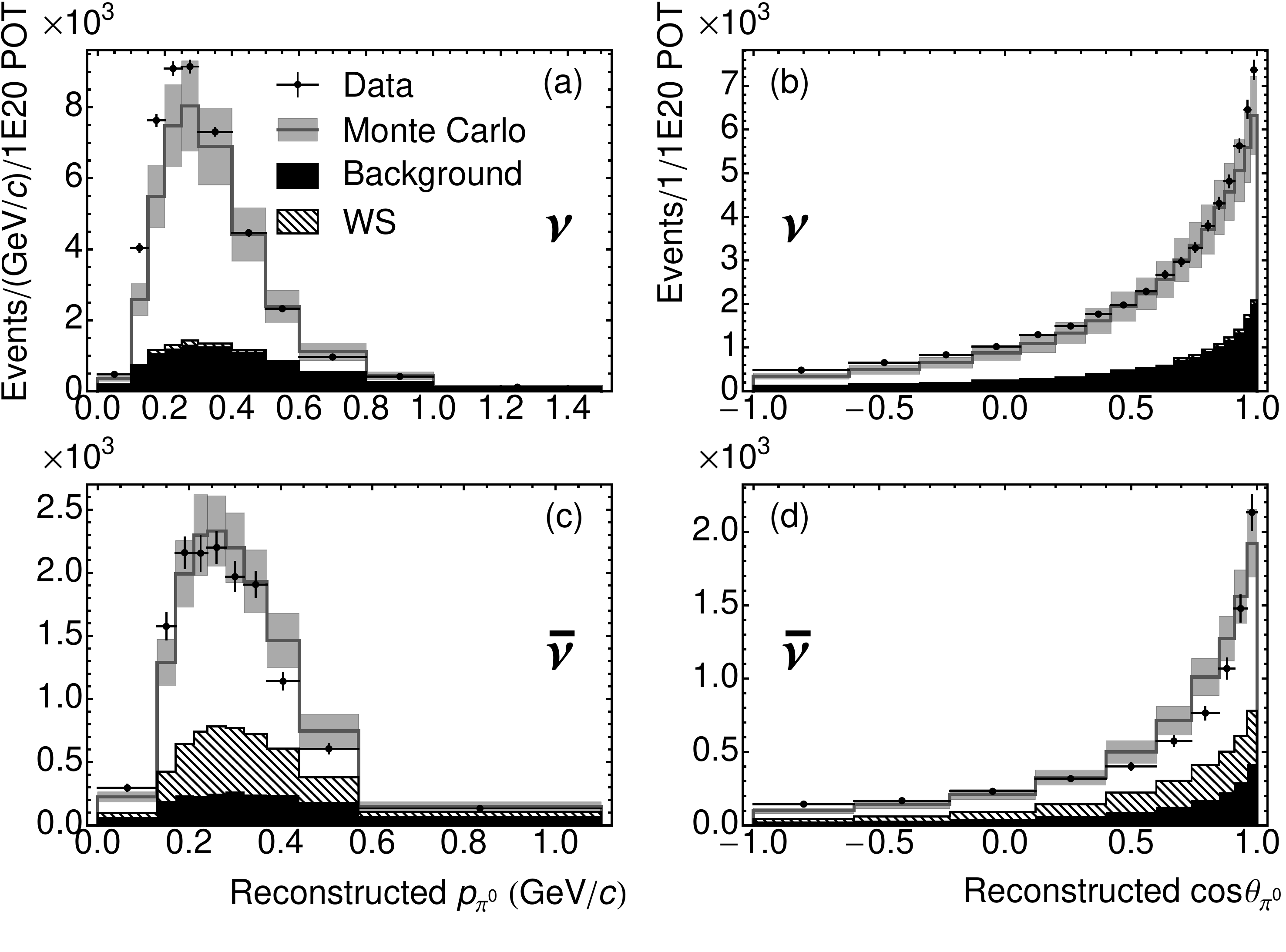}
\caption{\label{fig:boxdists}(a) The reconstructed $\pi^0$ momentum distribution for \ncpi{} candidates in neutrino mode running. The Monte Carlo distribution is shown as a dark-gray line and data as black dots. The box histogram is the systematic error on the Monte Carlo distribution; the error bars on the data are statistical only. Distributions are absolutely normalized to $10^{20}$ POT.  The black filled histogram is the non-\ncpi{} background and the hatched histogram above is the additional contribution from wrong-sign induced \ncpi{} production. (b) The reconstructed $\pi^0$ angle distribution in neutrino mode running.  (c) The reconstructed $\pi^0$ momentum distribution in antineutrino mode running. (d) The reconstructed $\pi^0$ angle distribution in antineutrino mode running.  }
\end{figure}

A selection of photon kinematic distributions from the $\pi^0$ fit appears in Fig.~\ref{fig:photonkinematics}.  An incorrect prediction of $\pi^0$s in the final state accounts for the disagreement between data and Monte Carlo in these distributions rather than any failure of the reconstruction, which has been separately vetted\cite{RECONSTRUCTION}.  Correcting the Monte Carlo with an \textit{in situ} measurement of the rate of $\pi^0$ production as a function of momentum---a kinematic that is strongly influenced by FSI---improves the level of agreement substantially\cite{PI0PLB}.  The photon kinematics are used to derive the $\pi^0$ kinematics.  The four-momentum of the $\pi^0$ is simply the sum-momentum of the two photons.  The incoming neutrino is assumed to be traveling in the beam direction, which is oriented with the $z$ axis by convention, so the $\pi^0$ angle is taken to be the angle relative to the $z$ axis.  Using the partitions appearing in Fig.~\ref{fig:boxdists}, we generate histograms of $\pi^0$ momentum and $\pi^0$ angle for the \ncpi{} candidates.  The neutrino mode $\pi^0$ momentum distribution extends to 1.5 $\text{GeV}/c$ while the antineutrino mode distribution extends to 1.1 $\text{GeV}/c$.

Background events arise from wrong-neutrino induced \ncpi{} production and interactions in the detector mimicking the signal signature.  Interactions occurring outside the detector (``dirt events'') introduce negligible background.  The fractional composition of the background is listed in Table~\ref{tab:background}.  Of the wrong-neutrino backgrounds, only \numu{}s in the \numubar{} beam constitute a significant background.  Indeed, because of the sizable contamination in the beam, wrong-sign production is the dominant background to the \numubar{} measurements; the \numu{} measurements are relatively unaffected by wrong-sign production.  The \nue{}(\nuebar{}) induced background is very small by virtue of the small beam contamination. The size of the detector affects the probability that particles emerging from the target nucleus will produce a $\pi^0$ in the tank. To avoid influencing the measurement with detector geometry, we include events with a $\pi^0$ produced anywhere outside the target nucleus (and no $\pi^0$ exiting the initial target nucleus) as background.  Background interactions typically mimic signal events through a combination of the production of a \piz{} outside the target nucleus and missed detection of other outgoing particles. NC $\pi^\pm$ production at the neutrino vertex is the most significant background to our signal.  The $\pi^\pm$ can readily charge exchange into a $\pi^0$.  NC elastic, multipion, CC $\pi^\pm$, and CC \piz{} interactions each contribute to the background at a similar level.  CC $\pi^\pm$ events mimic the signal in the same manner as their NC counterparts but also require that the outgoing lepton is undetected (captured or low momentum).  NC elastic events contribute via \piz{} production induced in the detector by the outgoing nucleon and multipion events through interactions producing a dominant \piz{}.  FSI creating additional mesons cause a small fraction of incoherent exclusive \ncpi{} events to be actually classified as background.
\begin{table}
\begin{ruledtabular}
\begin{tabular}{l@{\hspace{1em}}c@{\hspace{.5em}}c@{\hspace{1em}}l@{\hspace{1em}}c@{\hspace{.5em}}c}
Source&$ \bm{\nu}  $&$ \bar{\bm{\nu} } $&Source&$ \bm{\nu}  $&$ \bar{\bm{\nu} } $\\
\hline\\[-8pt]
NC $\pi^\pm$ & 23.0\% & 13.2\% & DIS & 3.5\% & 1.0\%\\
CC $\pi^\pm$ & 14.8\% & 4.5\% & CC QE & 5.0\% & 0.8\%\\
CC $\pi^0$ & 10.5\% & 3.5\% & $\text{K},\rho,\eta$ Prod. & 5.0\% & 2.5\%\\
Multi-$\pi$ & 12.8\% & 5.3\% & Other & 1.4\% & 2.1\%\\\cline{4-6}\\[-8pt]
NC Elastic & 12.4\% & 7.1\% & Wrong-Sign & 4.6\% & 56.1\%\\
NC $1\pi^0$ & 5.0\% & 2.5\% & $\nu_e+\bar{\nu}_e$ & 1.8\% & 1.4\%\\
\end{tabular}
\end{ruledtabular}
\caption{\label{tab:background}Predicted fractional composition of \ncpi{} background in neutrino and antineutrino modes broken down by exclusive channel at the initial neutrino interaction vertex and wrong-neutrino source.}
\end{table}

As the initial step to extract the cross section, the Monte Carlo prediction of the background rates is used to extract the signal rate from the \ncpi{} sample. We subtract the absolutely normalized rate of all backgrounds except the wrong-sign \ncpi{} background from the rate of candidate events in each bin of the kinematic distributions.  To remove the wrong-sign content, we multiply the remaining content of each bin by the estimate of the right-sign \ncpi{} fraction in that bin.

\begin{figure}
\includegraphics[width=\columnwidth]{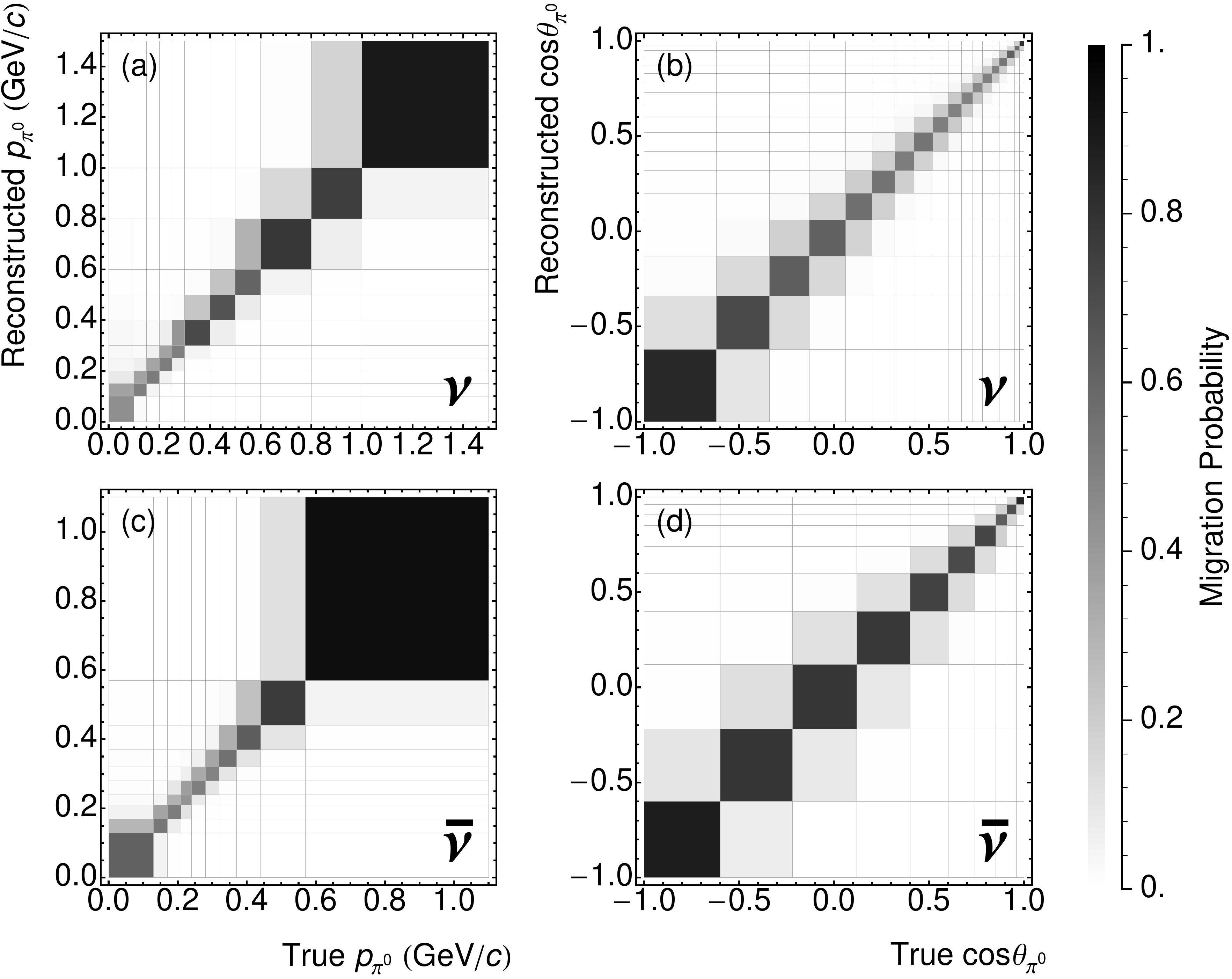}
\caption{\label{fig:smearingmatrices} (a) Response matrix for the measurement of $\pi^0$ momentum in \ncpi{} events satisfying selection cuts in neutrino mode. (b) The same for $\pi^0$ angle in neutrino mode. (c) The same for $\pi^0$ momentum in antineutrino mode.  (d) The same for $\pi^0$ angle in antineutrino mode.}
\end{figure}

Biases in the reconstruction, as well as detector effects, smear the measured kinematics of the outgoing pion.  This distortion is characterized in the \textit{response matrix}, $\bv{R}$.  For a measurement, $x$, and a partition of the domain of $x$, $(X_n)$, $R_{ij}$ is the probability that the reconstructed value of $x$ is in bin $i$ of $(X_n)$ if the true value of $x$ is in bin $j$.  The response matrices for our four measurements, as estimated by Monte Carlo, appear in Fig.~\ref{fig:smearingmatrices}.  The response matrices indicate a tendency of the reconstruction to slightly overestimate $\pi^0$ momentum, especially at low momentum.  In contrast, the response matrices for the measurement of $\pi^0$ angle demonstrate little bias and excellent resolution in the forward region.
In order to produce a physically meaningful measurement rather than one idiosyncratic to the experiment, we correct the measurement for this distortion using a process known as unsmearing (or unfolding).  Since the \numu{} and \numubar{} distributions differ in statistics by an order of magnitude and the \ppin{} and \costheta{} distributions differ radically in shape, using only one unsmearing technique is not necessarily appropriate.  We evaluate three options---applying (1) Tikhonov regularized unsmearing with the regularization strength chosen by the SVD prescription detailed by H{\"o}cker and Kartvelishvili\cite{SVDUNFOLD}, (2) a method analogous to one iteration of a Bayesian approach described by D'Agostini\cite{BAYESUNFOLD}, and (3)  no unsmearing---and select the least-biased result according to an unsmearing bias estimate from Cowan\cite{COWAN}.  The unsmearing methods are described in greater detail in Appendix~\ref{sec:unsmearing}.  We do not use matrix inversion to unsmear since it produces results with unacceptably large variance.  We apply method (1) to the \numu{} \ppin{} distribution, method (2) to the \numu{} \costheta{} and \numubar{} \ppin{} distributions, and method (3) to the \numubar{} \costheta{} distribution.

\begin{figure}
\includegraphics[width=\columnwidth]{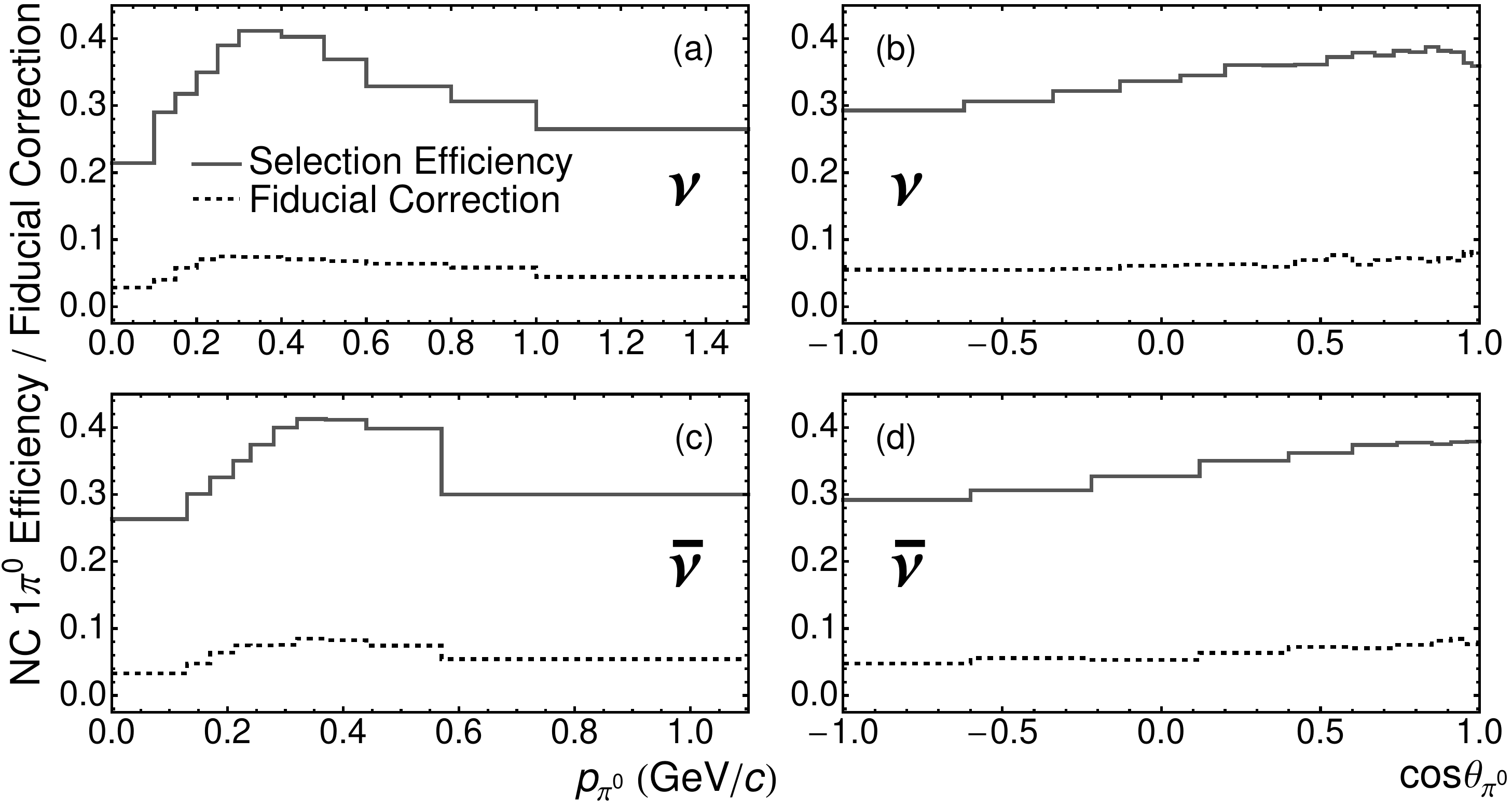}
\caption{\label{fig:efficiencies}(a) \ncpi{} selection efficiency (solid line) and fractional decrease in the number of events in the fiducial volume when using the true vertex versus the reconstructed vertex (dashed line) as functions of $\pi^0$ momentum in neutrino mode.  (b) The same for $\pi^0$ angle in neutrino mode. (c) The same for $\pi^0$ momentum in antineutrino mode.  (d) The same for $\pi^0$ angle in antineutrino mode.}
\end{figure}

After unsmearing the kinematic distributions, we apply corrections to compensate for the misestimation of the number of events in the fiducial volume due to misreconstructed interaction vertices and losses due to detection inefficiency.  These corrections appear in Fig.~\ref{fig:efficiencies}.  In the former case, a bias in the reconstruction to pull interaction vertices to the center of the detector leads to a 7\% excess of events being counted in the fiducial volume.  We subtract the fraction of nonfiducial events from each bin of each distribution.  The average \ncpi{} selection efficiency for each measurement is 36\%.  The selection efficiency is  momentum dependent: it is diminished at high and low momentum.  At low momentum, the $\log(\mathcal{L}_e/\mathcal{L}_\mu)$ cut becomes more inefficient as the ability of the reconstruction to discriminate between muonlike and electronlike events is reduced.  At higher momentum, loss of containment causes a larger proportion of signal events to fail the veto PMT hits requirement.  Loss of containment is responsible for the rejection of 11\% of signal events in neutrino mode and 13\% in antineutrino mode.  To recover the rate of events, we divide the kinematic distributions by the efficiency in each bin.

\begin{figure}
\includegraphics[width=\columnwidth]{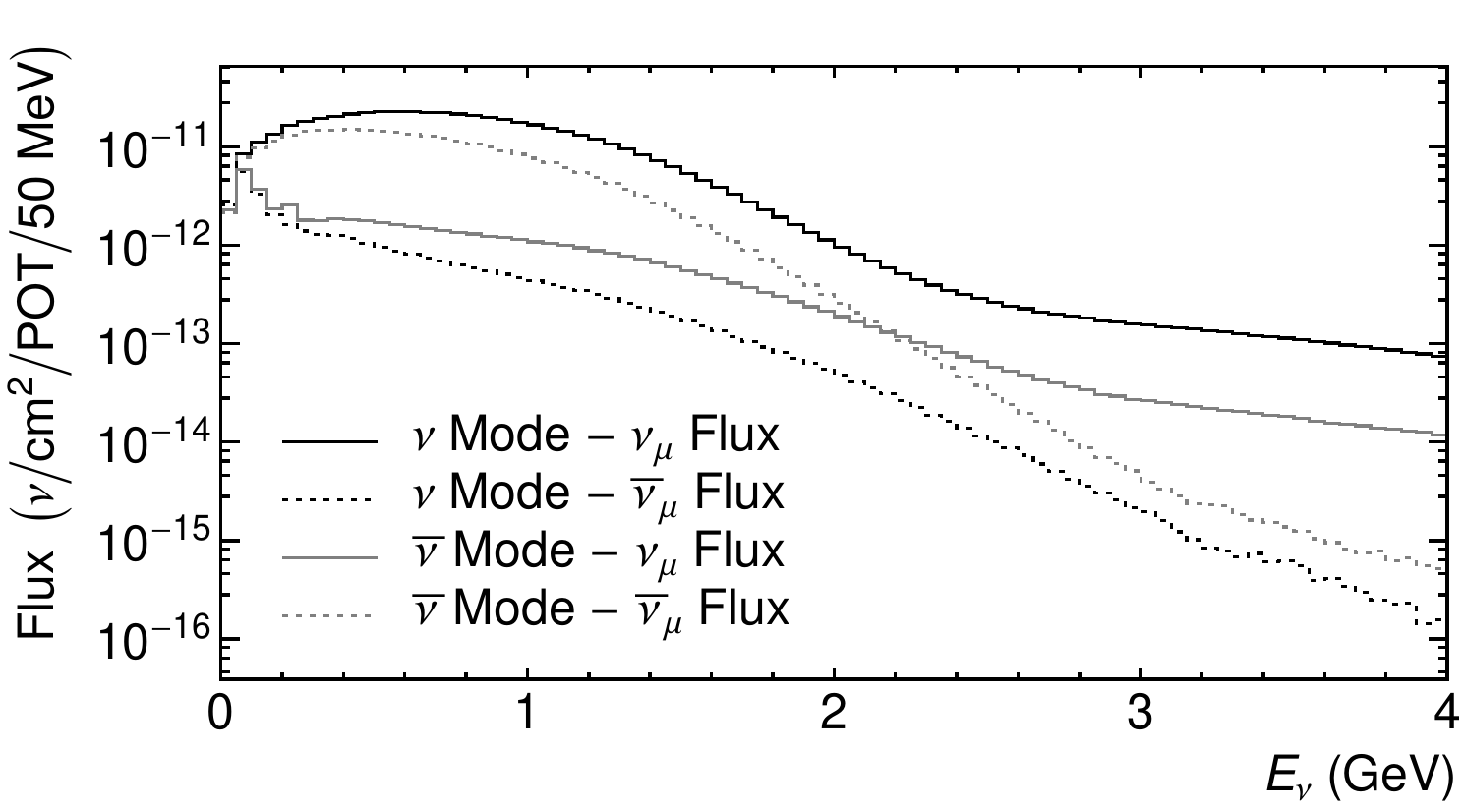}
\caption{\label{fig:fluxes}The predicted flux of \numu{} (solid lines) and \numubar{} (dotted lines) at the MiniBooNE detector with the horn configured in neutrino mode (black lines) and antineutrino mode (gray lines). The flux prediction is available at the MiniBooNE website\cite{FLUXRELEASE}.}
\end{figure}

With the rate of \ncpi{} production recovered, we must divide by the integrated flux and the number of targets to recover the flux-averaged cross section.  We predict the flux at MiniBooNE using a GEANT4-based simulation of the neutrino beam\cite{FLUXPAPER}.  Primary interactions of beam protons on the Be target producing $\pi^\pm$s, $\text{K}^{0,\pm}$s, protons, or neutrons are handled by a customized framework incorporating external data. In particular, the prediction of charged pion production (which is the dominant source of \numu{} and \numubar{}) is based on data from HARP\cite{HARP} and BNL E910\cite{E910}.  The flux prediction in both neutrino and antineutrino modes appears in Fig.~\ref{fig:fluxes}.  The simulation predicts an integrated flux of $(3.35\pm0.43_{sys})\times10^{11}\,\nu_\mu/\text{cm}^2$ over the course of neutrino mode running and $(1.08\pm0.12_{sys})\times10^{11}\,\bar\nu_\mu/\text{cm}^2$ over antineutrino mode running.  The uncertainty in the flux in neutrino (antineutrino) running can be split into 12.1\% (13.1\%) from secondary meson production uncertainties, 4.1\% (2.8\%) from the horn magnetic field (skin depth and current variations) and secondary interactions outside of the target, and 2\% (2\%) from the accounting of the number of protons delivered on target.  Using a measured value of $0.845\pm0.001\text{ g}/\text{cm}^2$ for the density of the mineral oil in the detector, we can determine that there are $2.664\pm0.003\times10^{32}$ nucleons in the 500 cm-radius fiducial volume.  Dividing each differential rate by the number of targets and the appropriate integrated flux yields the flux-averaged cross section per nucleon.

Plots of the resulting absolute differential cross sections for \ncpi{} production on $\text{CH}_2$ appear in Fig.~\ref{fig:xsecs} and the tables in Appendix~\ref{sec:xsectables}.  Per our signal definition, these cross sections include the effects of final state interactions. Integrating the differential cross sections yields total cross sections of $(4.76\pm0.05_{stat}\pm0.76_{sys})\times10^{-40}\text{ cm}^2/\text{nucleon}$ at a mean energy of $\langle E_\nu\rangle=808\text{ MeV}$ for \numu{}-induced production and $
(1.48\pm0.05_{stat}\pm0.23_{sys})\times10^{-40}\text{ cm}^2/\text{nucleon}$ at a mean energy of $\langle E_\nu\rangle=664\text{ MeV}$ for \numubar{}-induced production.  These cross sections are flux-averaged; hence, they are specific to the neutrino flux at MiniBooNE\cite{FLUXRELEASE}. Being the first absolute measurements of \ncpi{} production, there are no other measurements with which to compare.
\begin{figure*}
\includegraphics[width=\textwidth]{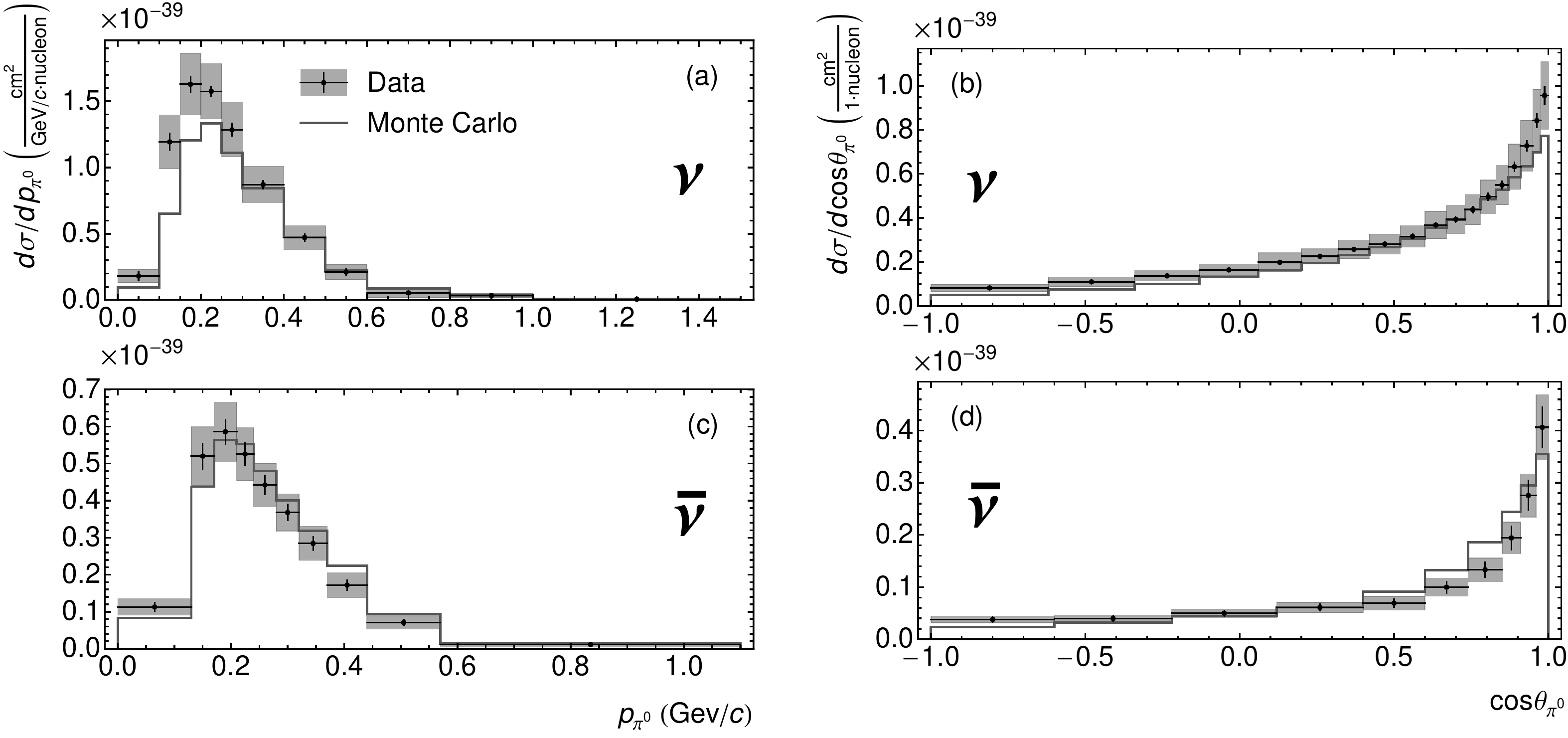}
\caption{\label{fig:xsecs}Flux-averaged absolute differential cross sections for \ncpi{} production on $\text{CH}_2$ including the effects of FSI.  Data are shown as black dots with statistical error bars and systematic error boxes.  The dark-gray line is the Monte Carlo prediction\cite{NUANCE} using R-S models of single pion production\cite{RSRES,RSCOH} modified as described in the text. (a) $\frac{d\sigma}{dp_{\pi^0}}$ for \numu{}-induced production. (b) $\frac{d\sigma}{d\cos\theta_{\pi^0}}$ for \numu{}-induced production. (c) $\frac{d\sigma}{dp_{\pi^0}}$ for \numubar{}-induced production. (d) $\frac{d\sigma}{d\cos\theta_{\pi^0}}$ for \numubar{}-induced production.  The numerical values for the cross sections appear in Appendix~\ref{sec:xsectables} and are also available at the MiniBooNE website\cite{DATARELEASE}.}
\end{figure*}
\section{\label{sec:systematics}Systematic Uncertainties}
Systematic uncertainties can be grouped into three principal categories---flux related, cross section related, and detector related. We gauge the uncertainty in the measurements including bin-to-bin correlations by calculating the covariance of the measurements over a set of Monte Carlo excursions wherein underlying parameters are varied within their uncertainties and correlations.

The same uncertainties affecting the integrated flux prediction detailed in Sec.~\ref{sec:analysis} also affect the Monte Carlo predictions used in the cross section calculation, \textit{e.g.} the background prediction.  In total, flux uncertainties produce a 12.4\% overall uncertainty in the \numu{} cross sections and 12.7\% in the \numubar{} cross sections.

The cross sections associated with background processes are varied within their uncertainties.  The relevant axial masses for quasielastic (QE), incoherent single pion, coherent single pion, and multipion production are varied by 6.2\%, 25\%, 27\%, and 40\% from their central values of 1.23 GeV/$c^2$, 1.10 GeV/$c^2$, 1.03 GeV/$c^2$, and 1.30 GeV/$c^2$, respectively.  The binding energy and Fermi momentum values used in the relativistic Fermi gas model\cite{RFG} underlying the simulation of QE, NC elastic, and incoherent pion production are varied by 26\% and 14\% from their central values of 34 MeV and 220 MeV/c, respectively.  The total normalization of QE scattering, deep inelastic scattering (DIS), and $\Delta$ radiative processes are varied by 10\%, 25\%, and 12.2\%, respectively. A Pauli blocking scale factor for CC QE events, $\kappa$\cite{CCQE}, is varied by 0.022 from its central value of 1.022.  In the target nucleus, the cross sections for pion absorption, pion charge exchange, and $\Delta$ interactions $(\Delta\text{N}\rightarrow\text{N}'\text{N})$, are varied by 25\%, 30\%, and 100\%, respectively.  Pion scattering cross sections in the mineral oil outside the target nucleus are varied by 35\% for absorption and 50\% for charge exchange.  The uncertainty in our pion interaction simulation is validated using external data for interactions on carbon\cite{PION1,PION2,PION3,PION4}.  In total, cross section uncertainties contribute an 8.4\% uncertainty in the measured \numu{} \ncpi{} production cross sections and 7.7\% in the \numubar{} cross sections.

Uncertainty in the optical model in the detector and PMT response as well as bias in the unsmearing make up the detector uncertainties.  Optical model uncertainties include variations in the amount of light production and in the propagation of light in the detector. A total of 39 parameters are varied.  For the PMT response, we assess one uncertainty by adjusting the discriminator threshold in the DAQ simulation from 0.1 PE to 0.2 PE and another by generating an excursion in the charge-time correlation of PMT hits.    We also assess the estimated bias in the unsmearing as an error.  Since unsmearing preserves the number of events in a distribution by design, the bias produces only a small uncertainty on the normalization of the cross section; the error is principally in the shape.  Detector uncertainties constitute a 5.1\% uncertainty in the \numu{} cross section and 4.8\% in the \numubar{} cross section. 

\section{\label{sec:discussion}Discussion}
\begin{figure}
\includegraphics[width=\columnwidth]{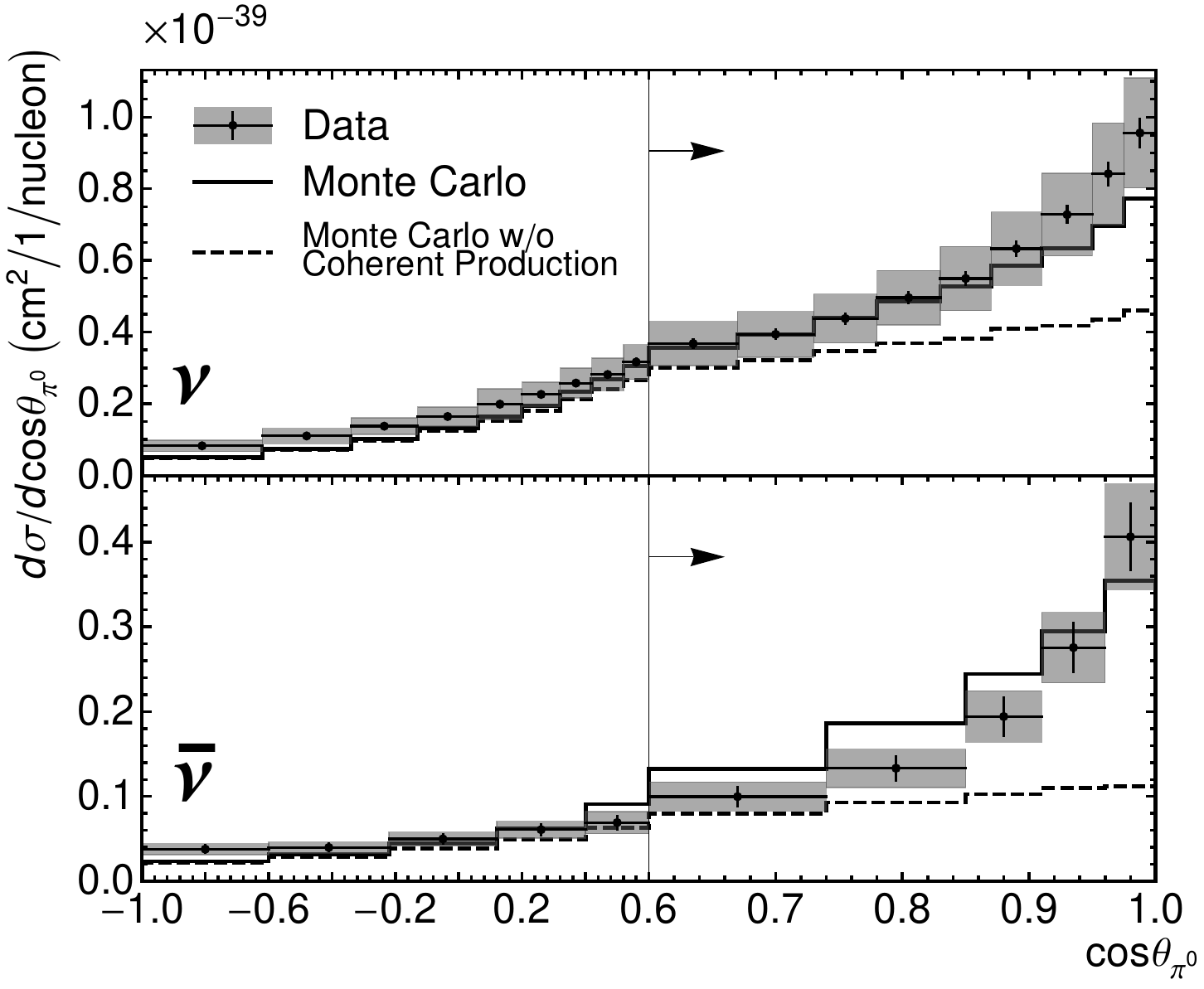}
\caption{\label{fig:cohcompplot}(a) The differential cross section for inclusive \ncpi{} production as a function of $\pi^0$ angle in neutrino mode (above) and antineutrino mode (below).  Data is indicated by black dots with statistical error bars and systematic error boxes.  The Monte Carlo prediction including the R-S single pion production models\cite{RSRES,RSCOH} as implemented in \textsc{Nuance} adjusted according to \cite{PI0PLB} is indicated by the thick black line.  The prediction omitting the coherent portion of NC $\pi^0$ production is indicated by the dashed black line.  The arrow indicates the region for which a $\chi^2$ is quoted in the text.  The horizontal scale is magnified in the forward region.}
\end{figure}

Honing models of single pion production continues to be of theoretical interest.  In particular, elucidating the nature of coherent pion production is a very active pursuit\cite{ALVAREZRUSO,AMAROPRD,HERNANDEZ,BERGER,PASCHOS,LEITNERCOH,NAKAMURA,GERSHTEIN,KOMACHENKO,MARTINI,KELKAR,BELKOV,SINGH}.  As an illustration, our own prediction of single $\pi^0$ production can be tested against our data.  

We predict single $\pi^0$ production using models by Rein and Sehgal\cite{RSRES,RSCOH} as implemented in \textsc{Nuance}. The axial masses for incoherent and coherent pion production are assumed to be $1.1\text{ GeV}/c^2$ and $1.03\text{ GeV}/c^2$, respectively.  Additionally, we use the \textsc{Nuance} FSI simulation in lieu of the pion absorptive factor suggested by R-S for coherent pion production.  Assuming these predictions\footnote{$\pi^0$ production was corrected in the Monte Carlo as a function of $\pi^0$ momentum using a proxy rate measurement from the data for the analysis in \cite{PI0PLB}}, MiniBooNE found that coherent pion production comprises $(19.5\pm1.1_{stat}\pm2.5_{sys})\%$ of exclusive NC $1\pi^0$ production in neutrino mode\cite{PI0PLB}.  This fraction implies a 35\% reduction in R-S coherent pion production (and a corresponding 5\% increase in incoherent production) that is incorporated into our Monte Carlo prediction.  Figure~\ref{fig:cohcompplot} compares the differential cross section in $\pi^0$ angle (the distribution most sensitive to the production mode) from data to our Monte Carlo prediction with and without coherent pion production. In the forward region above $\cos\theta_{\pi^0}=0.6$, the $\chi^2$ between neutrino (antineutrino) data and the Monte Carlo including coherent pion production is 8.23 (13.6) with 9 (5) degrees of freedom, which corresponds to a $p$-value of 0.511 (0.018).  Without coherent pion production, the $\chi^2$ worsens to 45.1 (25.7) with 9 (5) degrees of freedom, which corresponds to a $p$-value of $8.7\times10^{-7}$ (0.0001).  Both the neutrino and antineutrino data clearly favor the model of single $\pi^0$ production with nonzero coherent content.  Though the model including coherent pion production is favored, the shape disagreement evident in Fig.~\ref{fig:cohcompplot} substantiates, but does not confirm, the claims\cite{AMAROPRD,HERNANDEZ} that the R-S model\cite{RSCOH} is inadequate at neutrino energies below 2 GeV.  Alternative mechanisms, such as an incorrect prediction of the FSI\footnote{After correcting the predicted momentum dependence of $\pi^0$ production, MiniBooNE obtains good angular agreement with a reduced level of R-S\cite{RSCOH} coherent production in neutrino mode\cite{PI0PLB}. \emph{In this work, no momentum correction has been applied to the Monte Carlo}}, can account for the disagreement in part, but they are unlikely to explain the discrepancy in full, particularly in antineutrino mode.  Used in concert, our measurements in momentum and angle can be used to evaluate and refine the abundance of modern models that endeavor to correctly describe single pion production on nuclei with the effects of other mechanisms disentangled.

Our measurement is designed to be independent of the assumed models of single pion production and FSI.  Although, in making a pure \numu{} or \numubar{} measurement with a contaminated beam, we introduce some dependence on the assumed single pion production model by subtracting wrong-sign content.  In Appendix~\ref{sec:modeldependence}, we characterize this sensitivity and present an alternative, fully-independent measurement.

\begin{figure}
\includegraphics[width=\columnwidth]{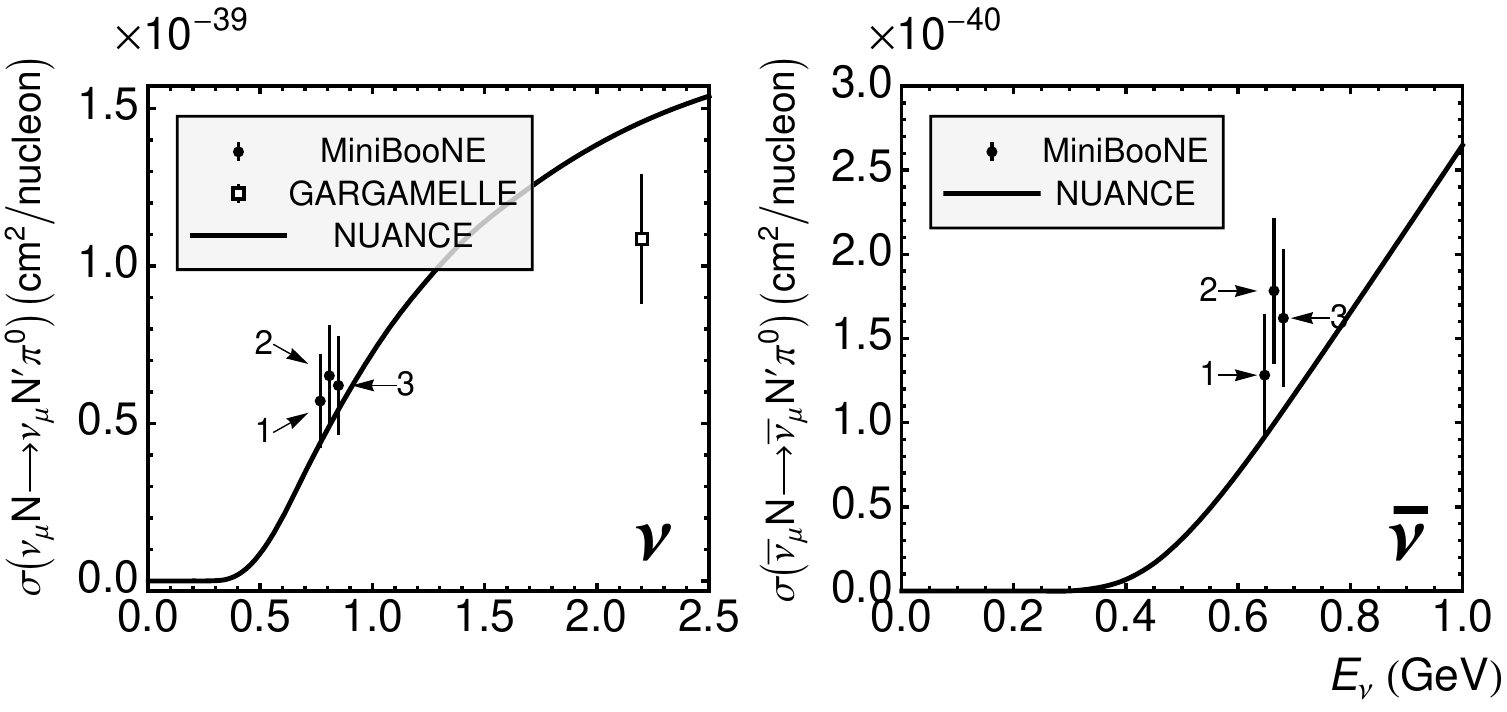}
\caption{\label{fig:resxsecs}(a) The flux-averaged total cross sections for \numu{}-induced incoherent exclusive \ncpi{} production on $\text{CH}_2$ corrected for FSI. Points 1, 2, and 3, are the cross sections extracted using the MiniBooNE implementation of the R-S model for coherent pion production, the model in \cite{AMAROPRD}, and the model in \cite{ALVAREZRUSO}, respectively.  The points are placed at the mean energy of the beam in neutrino mode; the spread is only for clarity.  The curve is the \textsc{Nuance} prediction using the R-S model. Also shown for comparison is the measurement made from the Gargamelle data\cite{GGMHAWKER}.  The Gargamelle experiment used a propane and freon $(\text{C}_3\text{H}_8 + \text{C}\text{F}_3\text{Br})$ target.  (b) The same for \numubar{}-induced incoherent exclusive \ncpi{} production.  In this case, there are no external measurements to compare to.}
\end{figure}

In addition, we assess the cross section for \numu{} and \numubar{} induced incoherent \ncpi{} production defined at the initial neutrino interaction vertex as a means to compare with past measurements. Such an exclusive measurement is naturally quite sensitive to assumed models of both single pion production and FSI.  We use the same selection cuts as in the primary analysis.  Because coherent \ncpi{} production is a background to this measurement, the result suffers from a fairly low predicted signal fraction: 57\% in neutrino mode and 34\% in antineutrino mode.  We use the same selection of unsmearing techniques used in the primary analysis as well.  The nonfiducial fraction is also predicted to be the same at 7\%.  Unlike in the inclusive measurement, the efficiency correction includes a correction for FSI predicted using Monte Carlo that recovers the kinematic distributions at the initial neutrino interaction vertex.  This overall efficiency including selection inefficiency and FSI is predicted to be 24\% in both neutrino and antineutrino modes.  After all corrections, we find the cross section to be $(5.71\pm0.08_{stat}\pm1.45_{sys})\times10^{-40}\text{ cm}^2/\text{nucleon}$ for \numu{}-induced incoherent exclusive NC $1\pi^0$ production on $\text{CH}_2$ and $(1.28\pm0.07_{stat}\pm0.35_{sys})\times10^{-40}\text{ cm}^2/\text{nucleon}$ for \numubar{}-induced production.  These cross sections are averaged over the MiniBooNE flux as well.  Here, the significance of FSI becomes apparent: the \numu{} incoherent exclusive \ncpi{} production cross section actually exceeds the \numu{} inclusive \ncpi{} production cross section.  Repeating the measurement using the models of \cite{AMAROPRD} and \cite{ALVAREZRUSO} discussed in Appendix~\ref{sec:modeldependence} yields values of $(6.51\pm0.08_{stat}\pm1.56_{sys})\times10^{-40}\text{ cm}^2/\text{nucleon}$ and $(6.20\pm0.08_{stat}\pm1.52_{sys})\times10^{-40}\text{ cm}^2/\text{nucleon}$, respectively, for \numu{} induced production, and $(1.78\pm0.07_{stat}\pm0.42_{sys})\times10^{-40}\text{ cm}^2/\text{nucleon}$ and $(1.62\pm0.07_{stat}\pm0.39_{sys})\times10^{-40}\text{ cm}^2/\text{nucleon}$, respectively, for \numubar{} induced production.  The variation in the measurements extracted under alternative models of coherent pion production illustrate the model dependence of the extracted incoherent cross section.  These measurements are plotted against prior measurements and the \textsc{Nuance} prediction (using R-S) in Fig.~\ref{fig:resxsecs}.  A comparison can be made only to the result of the reanalysis of the Gargamelle data\cite{GGMHAWKER} since the measurement at Aachen-Padova was limited to production on protons\cite{FAISSNER}.

\section{\label{sec:conclusion}Conclusion}
In conclusion, we have used the largest sample of NC $1\pi^0$ events collected to date to produce measurements of absolute differential cross sections of \ncpi{} production induced by both neutrinos and antineutrinos on $\text{CH}_2$ as functions of both \piz{} momentum and \piz{} angle averaged over the MiniBooNE flux.  These measurements, which are the principal result of this work, can be found in Fig.~\ref{fig:xsecs} and Table~\ref{tab:xsectables}.  The total cross sections have been measured to be $(4.76\pm0.05_{stat}\pm0.76_{sys})\times10^{-40}\text{ cm}^2/\text{nucleon}$ for \numu{} interactions at a mean energy of 808 MeV and $(1.47\pm0.05_{stat}\pm0.23_{sys})\times10^{-40}\text{ cm}^2/\text{nucleon}$ for \numubar{} interactions at a mean energy of 664 MeV.  These measurements should prove useful to both future oscillation experiments seeking to constrain their backgrounds and those developing models of single pion production seeking to test their predictions.  We have additionally measured total cross sections for incoherent exclusive \ncpi{} production on $\text{CH}_2$ to compare to a prior measurement.  These cross sections were found to be $(5.71\pm0.08_{stat}\pm1.45_{sys})\times10^{-40}\text{ cm}^2/\text{nucleon}$ for \numu{}-induced production and $(1.28\pm0.07_{stat}\pm0.35_{sys})\times10^{-40}\text{ cm}^2/\text{nucleon}$ \numubar{}-induced production.

\begin{acknowledgments}
We wish to acknowledge the support of Fermilab, the National Science Foundation, and the Department of Energy in the construction, operation, and data analysis of the MiniBooNE experiment.
\end{acknowledgments}

\appendix
\section{\label{sec:modeldependence}Measurement Model Dependence}
\begin{figure}
\includegraphics[width=\columnwidth]{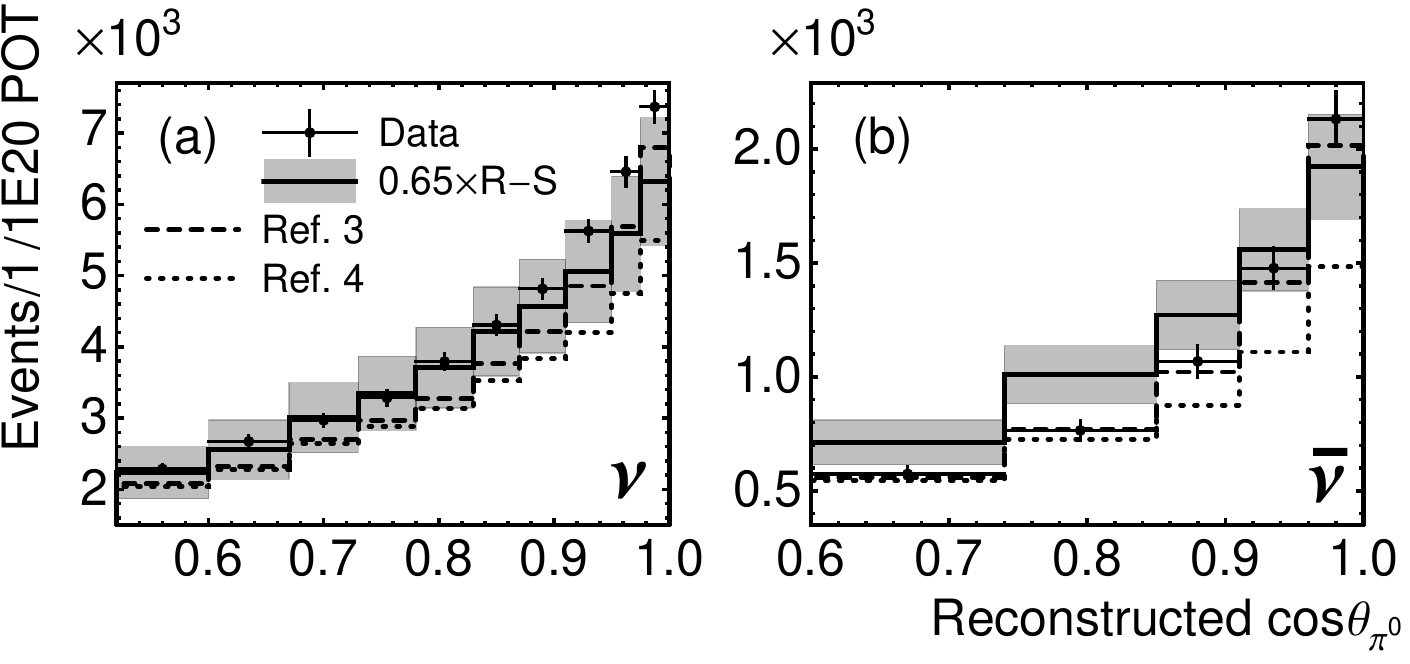}
\caption{\label{fig:modelcutdists}Pion angular distributions in the forward region for \ncpi{} candidates in (a) neutrino and (b) antineutrino mode running. Data is indicated by black dots with statistical error bars.  The Monte Carlo prediction using the rescaled\cite{PI0PLB} R-S model of coherent pion production as implemented in \textsc{Nuance}\cite{RSCOH,NUANCE} is indicated by the solid black line with gray systematic error boxes.  The predictions using the models of \cite{AMAROPRD} and \cite{ALVAREZRUSO} are indicated by the dotted line and the dashed line, respectively. The systematic error in the predictions using the alternative models is of the same relative size as the prediction using R-S; it is omitted for clarity. Distributions are normalized to $10^{20}$ POT (b) The same for antineutrino mode.}
\end{figure}
\begin{figure}
\includegraphics[width=\columnwidth]{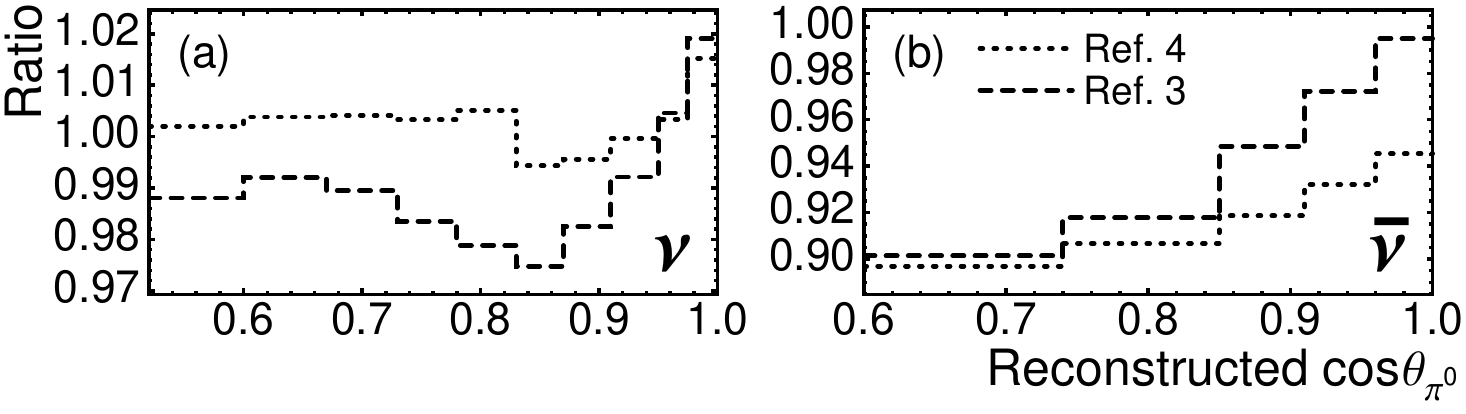}
\caption{\label{fig:modelxsecratio}(a) The ratio of the \numu{} \ncpi{} angular cross section extracted using the coherent production models in \cite{AMAROPRD} (dotted line) and \cite{ALVAREZRUSO} (dashed line) compared to the principal result extracted using the R-S model. (b) The same for the \numubar{} \ncpi{} cross section.}
\end{figure}

Subtraction of wrong-sign induced \ncpi{} signal events inevitably couples our measurements to the assumed model of \ncpi{} production.  For the sake of example, we considered the effect of substituting the coherent pion production models of Refs.~\cite{AMAROPRD}~\&~\cite{ALVAREZRUSO} into our Monte Carlo prediction.  The difference in the angular distribution of events satisfying the \ncpi{} selection cuts under these models appears in Fig.~\ref{fig:modelcutdists}.  Both the microscopic models demonstrate a sharper peaking in forward direction compared to the MiniBooNE R-S central value.  However, owing to a different choice for the N--$\Delta$ transition axial form factor $C^A_5$, Ref.~\cite{AMAROPRD} predicts substantially less production than Ref.~\cite{ALVAREZRUSO}.  In Fig.~\ref{fig:modelxsecratio}, the ratio of the angular cross sections extracted assuming the models in Refs.~\cite{AMAROPRD,ALVAREZRUSO} relative to the primary result is shown.  Because of the low wrong-sign contamination, the \numu{} cross section is relatively insensitive to changes in the model; however the \numubar{} cross section deviates more significantly under the model variations.  The \numubar{} total cross section decreases by 5.8\% under \cite{AMAROPRD} and 4.4\% under \cite{ALVAREZRUSO}; the \numu{} total cross section varies by $<1\%$ in either case.  Even though an attempt is made to partially mitigate model dependence in the wrong-sign subtraction by scaling by the right-sign fraction rather than outright subtracting the rate, the large wrong-sign fraction in antineutrino mode together with the very large variation from \cite{AMAROPRD} conspire to generate a non-negligible difference in the measured cross section.  Such dependence is unavoidable when measuring a \numubar{}-only cross section.  

\begin{figure}
\includegraphics[width=\columnwidth]{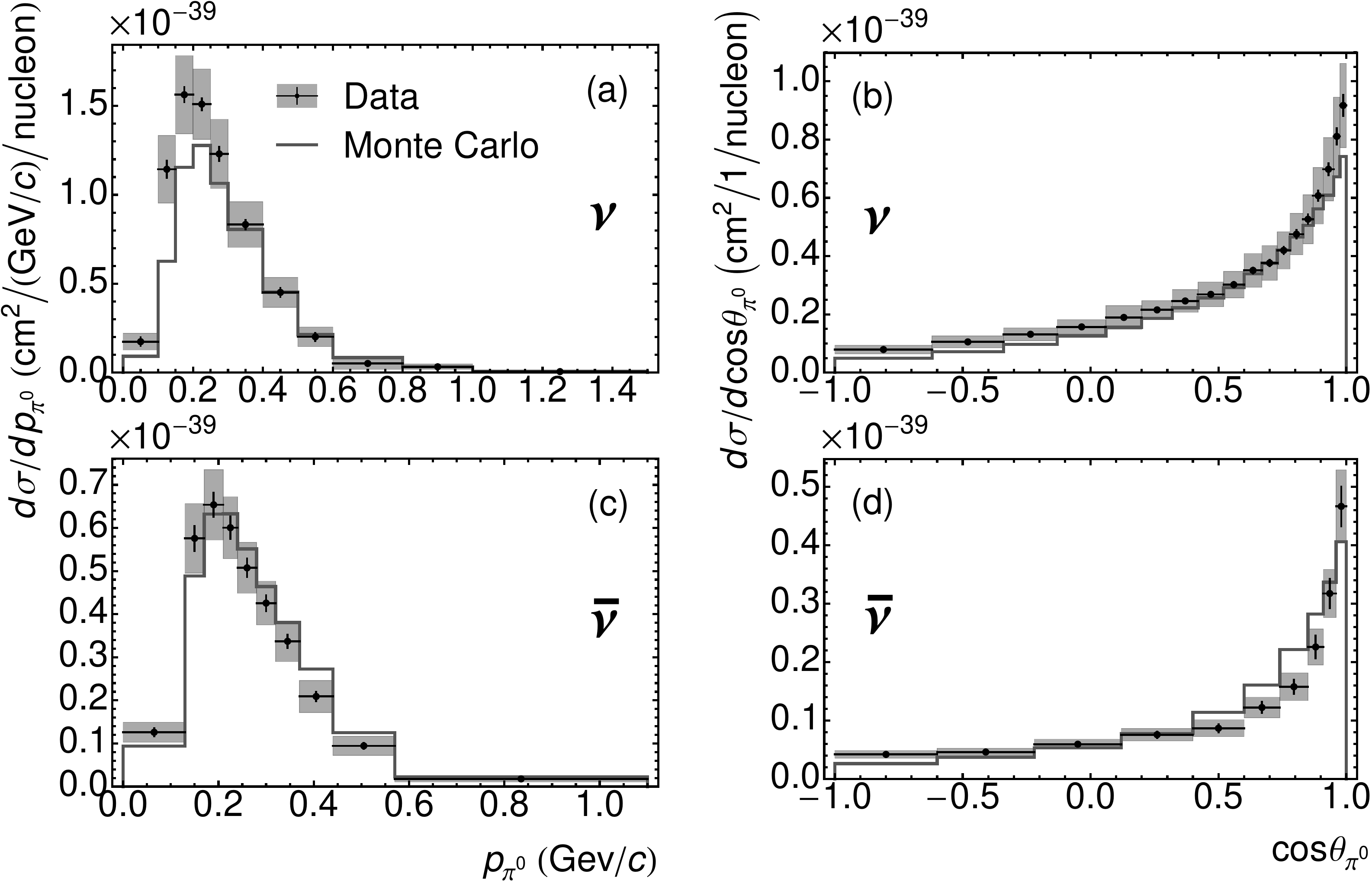}
\caption{\label{fig:asxsecs}Flux-averaged absolute differential cross sections for \numu{}+\numubar{}-induced \ncpi{} production on $\text{CH}_2$.  Data are shown as black dots with statistical error bars and systematic error boxes.  The dark-gray line is the Monte Carlo\cite{NUANCE} prediction using R-S models of single pion production\cite{RSRES,RSCOH} modified as described in the text. (a) $\frac{d\sigma}{dp_{\pi^0}}$ in neutrino mode. (b) $\frac{d\sigma}{d\cos\theta_{\pi^0}}$ in neutrino mode. (c) $\frac{d\sigma}{dp_{\pi^0}}$ in antineutrino mode. (d) $\frac{d\sigma}{d\cos\theta_{\pi^0}}$ in antineutrino mode.}
\end{figure}

In order to provide a measurement that is unbiased by any assumed model of \ncpi{} production, against which other models can be tested, we performed the principal analysis again in the exact same manner except signal events induced by wrong-sign neutrinos are not subtracted.  These combined \numu{}+\numubar{} measurements are almost entirely free of the model dependence introduced by the wrong-sign subtraction at the cost of being a less immediately meaningful measurement. Naturally, the signal fraction increases: it is 75\% in neutrino mode and 82\% in antineutrino mode.  The nonfiducial fraction and selection efficiency remain the same (7\% and 36\%, respectively).  The combined integrated flux over neutrino mode running is  $(3.57\pm0.50_{sys})\times10^{11}\,(\nu_\mu+\bar\nu_\mu)/\text{cm}^2$ and the combined integrated flux over antineutrino mode running is $(1.58\pm0.21_{sys})\times10^{11}\,(\nu_\mu+\bar\nu_\mu)/\text{cm}^2$.  We find the flux-averaged total cross section for \numu{}+\numubar{}-induced \ncpi{} production on $\text{CH}_2$ to be $(4.56\pm0.05_{stat}\pm0.71_{sys})\times10^{-40}\text{ cm}^2/\text{nucleon}$ in neutrino mode and $(1.75\pm0.04_{stat}\pm0.24_{sys})\times10^{-40}\text{ cm}^2/\text{nucleon}$ in antineutrino mode.  The \numu{}+\numubar{} differential cross sections appear in Fig.~\ref{fig:asxsecs}.

\section{\label{sec:unsmearing}Unsmearing}
We begin by defining an abstract unsmearing scenario.  Suppose we make a measurement of a variable $x$ over an $n$-bin partition of the domain of $x$, $(X_n)$, that is subject to smearing dictated by a response matrix $\bv{R}$.  If the discrete probability density function for $x$ over the partition is $\bm{\alpha}$, then the probability distribution function for the measured values is $\bm{\beta}=\bv{R}\bm{\alpha}$.  In an actual measurement of $N$ events, we make a draw $\bv{b}\sim N\bm{\beta}$ which corresponds to an unknown true distribution $\bv{a}\sim N\bm{\alpha}$.  In unsmearing, we seek to determine an estimator for $\bv{a}$, $\hat{\bv{a}}$, knowing only $\bv{b}$ and Monte Carlo estimates of $\bv{R}$ and $\bm{\alpha}$, $\bv{R}^{MC}$ and $\bm{\alpha}^{MC}$.  In this analysis, we treat smearing as affecting only the shape of a distribution and not the normalization as including efficiency losses would do.  Here we describe three unsmearing methods, two of which are used in the analysis.

A naive method of unsmearing follows from the expression $\bm{\beta}=\bv{R}\bm{\alpha}$ given the population distributions and the response matrix.  It follows that $\bm{\alpha}=\bv{R}^{-1}\bm{\beta}$.  Hence, if $\bv{R}^{MC}$ estimates $\bv{R}$ well, then we may choose 
\begin{equation}
  \label{eq:matrixinv}
  \bv{\hat{a}} = {\bv{R}^{MC}}^{-1}\bv{b}
\end{equation}
to be an estimator for $\bv{a}$.  This choice of unsmearing is knows as matrix inversion.  Since Eq.~\ref{eq:matrixinv} involves the inversion of a matrix, it is particularly sensitive to perturbations in $\bv{R}^{MC}$ and $\bv{b}$.  Matrix inversion often proves to be too unstable to be useful.

The second method is a specialization of Tikhonov regularization.  Under Tikhonov regularization we choose the $\hat{\bv{a}}$ that minimizes the quantity
\begin{equation}
  \label{eq:tr}
  (\bv{R}^{MC}\hat{\bv{a}}-\bv{b})^{\textsf{T}}\bv{V}\left(\bv{b}\right)(\bv{R}^{MC}\hat{\bv{a}}-\bv{b}) + \tau\|\bv{L}\hat{\bv{a}}\|^2,
\end{equation}
where $\bv{V}\left(\bv{b}\right)$ is the covariance matrix for $\bv{b}$, $\bv{L}$ is some linear operator, and $\tau$ is a constant controlling the strength of regularization.  The quantity on the left is simply a $\chi^2$ between the measured reconstructed distribution and the smeared estimator for the true distribution.  Minimizing only the $\chi^2$ results in the estimator $\hat{\bv{a}} = {\bv{R}^{MC}}^{-1}\bv{b}$---the result of matrix inversion.  This result is usually highly unstable.  The right hand term is a regularizing term that reduces the variance by adding a penalty for not satisfying some \textit{a priori} characteristic of $\hat{\bv{a}}$ encoded by the action of $\bv{L}$.  For this analysis, we assume that the true distributions are smooth, so we seek to minimize the curvature of the estimate.  To that end, we choose $\bv{L}$ to be the second finite-difference operator (a discretization of the second derivative).  Equation \ref{eq:tr} can by minimized analytically.  Typically no constraint is placed on the minimization, but we use the method of Lagrange multipliers to minimize under the constraint that $\sum_i \hat{a}_i = \sum_i b_i$ per our objective to not change the normalization, which results in
\begin{eqnarray}
\hat{\bv{a}} &=& \bv{U}'\bv{b}+\left(\sum_{ij} (\delta_{ij}-U'_{ij})v_j\right)\bv{s},\notag\\
\bv{U}'&\equiv& \left(\bv{R}^{MC}+\tau \bv{V} {\bv{R}^{MC}}^{\textsf{T}^{-1}}\bv{L}^{\textsf{T}}\bv{L}\right)^{-1},\notag\\
s_i&\equiv& \frac{\sum_j U'_{ik}V_{kl}{R^{MC}}^{-1}_{jl}}{\sum_{jk} U'_{jl}V_{lm}{R^{MC}}^{-1}_{km}}.
\end{eqnarray}
The choice of $\tau$ follows the prescription in Ref.~\cite{SVDUNFOLD}.  Bias is introduced not through the Monte Carlo, but the choice of Tikhonov matrix, $\bv{L}$.

The third method is equivalent to a single iteration of the Bayesian method described in Ref.~\cite{BAYESUNFOLD}.  Since $\sum_j S_{ji}^{MC} = 1 ~\forall~i$ by definition, it follows that ${\bv{S}^{MC}}^{\textsf{T}}\cdot(1,1,...,1)=(1,1,...,1)$.  We construct a matrix $\bv{U}$, given by
\begin{equation}
  \bv{U}\equiv \mathop{\rm diag}(\bv{\alpha}^{MC}){\bv{S}^{MC}}^{\textsf{T}}\mathop{\rm diag}(\bv{\beta}^{MC})^{-1}.
\end{equation}
By construction $\bv{U}\bv{\beta}^{MC} = \bv{\alpha}^{MC}$.  Assuming that the Monte Carlo is a good estimator for the data, then we can use $\bv{\hat{a}} = \bv{U}\bv{b}$ as an estimator for $\bv{a}$.  This method introduces bias from the Monte Carlo.

\section{\label{sec:xsectables}Cross Section Values}
The \numu{}- and \numubar-induced \ncpi{} production cross section measurements on $\text{CH}_2$ are tabulated in Table~\ref{tab:xsectables}. The measurements together with full error matrices are also in a data release available at the MiniBooNE website\cite{DATARELEASE}.

\captionsetup[subfloat]{singlelinecheck=false,justification=RaggedRight,margin=2pt}

\begin{table}
\scriptsize
\subfloat[\numu{} \ncpi{} production \ppin{} differential cross section]{
\begin{ruledtabular}
\begin{tabular}{cccc}
$p_{\pi^0}$ $(\text{GeV}/c)$&\parbox[c]{9em}{$d\sigma/d p_{\pi^0}$\\{\tiny $(10^{-39}\text{cm}^2/(\text{GeV}/c))$}}&$p_{\pi^0}$ $(\text{GeV}/c)$&\parbox[c]{9em}{$d\sigma/d p_{\pi^0}$\\{\tiny$(10^{-39}\text{cm}^2/(\text{GeV}/c))$}}\\[6pt]
\hline\\[-6pt]
$(0.00,0.10)$ & $0.18\pm0.06$ & $(0.40,0.50)$ & $0.47\pm0.09$\\
$(0.10,0.15)$ & $1.19\pm0.21$ & $(0.50,0.60)$ & $0.21\pm0.06$\\
$(0.15,0.20)$ & $1.63\pm0.24$ & $(0.60,0.80)$ & $0.05\pm0.04$\\
$(0.20,0.25)$ & $1.58\pm0.21$ & $(0.80,1.00)$ & $0.03\pm0.02$\\
$(0.25,0.30)$ & $1.28\pm0.21$ & $(1.00,1.50)$ & $0.01\pm0.01$\\
$(0.30,0.40)$ & $0.87\pm0.14$ &  & \\
\end{tabular}
\end{ruledtabular}}\\
\subfloat[\numu{} \ncpi{} production \costheta{} differential cross section]{
\begin{ruledtabular}
\begin{tabular}{cccc}
$\cos\theta_{\pi^0}$&\parbox[c]{6em}{$d\sigma/d\cos\theta_{\pi^0}$\\$(10^{-40}\text{cm}^2/1)$}&$\cos\theta_{\pi^0}$&\parbox[c]{6em}{$d\sigma/d\cos\theta_{\pi^0}$\\$(10^{-40}\text{cm}^2/1)$}\\[6pt]
\hline\\[-6pt]
$(-1.000,-0.620)$ & $0.82\pm0.16$ & $(+0.600,+0.670)$ & $3.68\pm0.63$\\
$(-0.620,-0.340)$ & $1.10\pm0.22$ & $(+0.670,+0.730)$ & $3.94\pm0.65$\\
$(-0.340,-0.130)$ & $1.37\pm0.24$ & $(+0.730,+0.780)$ & $4.38\pm0.70$\\
$(-0.130,+0.060)$ & $1.64\pm0.27$ & $(+0.780,+0.830)$ & $4.96\pm0.77$\\
$(+0.060,+0.200)$ & $1.99\pm0.43$ & $(+0.830,+0.870)$ & $5.49\pm0.91$\\
$(+0.200,+0.320)$ & $2.26\pm0.35$ & $(+0.870,+0.910)$ & $6.33\pm1.04$\\
$(+0.320,+0.420)$ & $2.58\pm0.43$ & $(+0.910,+0.950)$ & $7.28\pm1.18$\\
$(+0.420,+0.520)$ & $2.82\pm0.46$ & $(+0.950,+0.975)$ & $8.42\pm1.45$\\
$(+0.520,+0.600)$ & $3.16\pm0.50$ & $(+0.975,+1.000)$ & $9.56\pm1.58$\\
\end{tabular}
\end{ruledtabular}}\
\subfloat[\numubar{} \ncpi{} production \ppin{} differential cross section]{
\begin{ruledtabular}
\begin{tabular}{cccc}
$p_{\pi^0}$ $(\text{GeV}/c)$&\parbox[c]{9em}{$d\sigma/d p_{\pi^0}$\\{\tiny $(10^{-40}\text{cm}^2/(\text{GeV}/c))$}}&$p_{\pi^0}$ $(\text{GeV}/c)$&\parbox[c]{9em}{$d\sigma/d p_{\pi^0}$\\{\tiny$(10^{-40}\text{cm}^2/(\text{GeV}/c))$}}\\[6pt]
\hline\\[-6pt]
$(0.00,0.13)$ & $1.13\pm0.25$ & $(0.28,0.32)$ & $3.68\pm0.55$\\
$(0.13,0.17)$ & $5.20\pm0.86$ & $(0.32,0.37)$ & $2.84\pm0.49$\\
$(0.17,0.21)$ & $5.86\pm0.86$ & $(0.37,0.44)$ & $1.72\pm0.36$\\
$(0.21,0.24)$ & $5.26\pm0.78$ & $(0.44,0.57)$ & $0.71\pm0.19$\\
$(0.24,0.28)$ & $4.42\pm0.64$ & $(0.57,1.10)$ & $0.11\pm0.06$\\
\end{tabular}
\end{ruledtabular}}\\
\subfloat[\numubar{} \ncpi{} production \costheta{} differential cross section]{
\begin{ruledtabular}
\begin{tabular}{cccc}
$\cos\theta_{\pi^0}$&\parbox[c]{6em}{$d\sigma/d\cos\theta_{\pi^0}$\\$(10^{-40}\text{cm}^2/1)$}&$\cos\theta_{\pi^0}$&\parbox[c]{6em}{$d\sigma/d\cos\theta_{\pi^0}$\\$(10^{-40}\text{cm}^2/1)$}\\[6pt]
\hline\\[-6pt]
$(-1.00,-0.60)$ & $0.38\pm0.08$ & $(+0.60,+0.74)$ & $1.00\pm0.20$\\
$(-0.60,-0.22)$ & $0.40\pm0.08$ & $(+0.74,+0.85)$ & $1.33\pm0.27$\\
$(-0.22,+0.12)$ & $0.50\pm0.10$ & $(+0.85,+0.91)$ & $1.94\pm0.38$\\
$(+0.12,+0.40)$ & $0.61\pm0.12$ & $(+0.91,+0.96)$ & $2.76\pm0.50$\\
$(+0.40,+0.60)$ & $0.69\pm0.15$ & $(+0.96,+1.00)$ & $4.06\pm0.74$\\
\end{tabular}
\end{ruledtabular}}
\caption{\label{tab:xsectables}Tabulated values of the flux-averaged differential cross sections for \numu{}- and \numubar{}-induced \ncpi{} production on $\text{CH}_2$ corresponding to the plots in Figure~\ref{fig:xsecs}.  The error quoted with the cross section values is the quadrature sum of the diagonal statistical and systematic error.}
\end{table}

\bibliographystyle{apsrev}
%\bibliography{bibfile}

\end{document}